\documentclass[]{article}
\usepackage[margin=1in]{geometry}
\usepackage[utf8]{inputenc}
\usepackage[english]{babel}
\usepackage{graphicx,amssymb,amsmath,amsfonts,bm,array,slashed,mathtools,multirow,float}
\usepackage{subcaption}
\usepackage{booktabs} 
\usepackage[colorlinks=true,linkcolor=blue,urlcolor=blue,citecolor=blue]{hyperref}
\DeclareMathOperator{\Tr}{Tr}
\title{Open charm mesons in variational scheme and HQET}
\author{K. K. Vishwakarma$^{a}$\thanks{ vish.kumar.kundan@gmail.com (corresponding authors)}, Ritu Garg$^b$\thanks{ritugarg039@gmail.com (corresponding authors)}, Alka Upadhyay$^a$\thanks{alka@thapar.edu}}
 \date{$^a$Thapar Institute of Engineering and Technology, Patiala, India-147004 \\
 $^b$Department of Physics, Manipal University, Jaipur, India-303007}
\DeclareUnicodeCharacter{2217}{*}

\begin{document}
\maketitle
\begin{abstract} 
The charm ($D$) and charm-strange ($D_s$) mesons are investigated in a variational scheme using  Gaussian trial wave functions. The Hamiltonian contains Song and Lin potential with a constant term dependent on radial and orbital quantum numbers. The Gaussian wave function used has a dependence on radial distance $r$, radial quantum number $n$, orbital quantum number $l$ and a trial parameter $\mu$. 
The obtained spectra of $D$ and $D_s$ mesons are in good agreement with other theoretical models and available experimental masses. The mass spectra of $D$ and $D_s$ mesons are also used to plot Regge trajectories in the ($J$, $M^2$) and ($n_r$, $M^2$) planes. In ($J$, $M^2$) plane, both natural and unnatural parity states of $D$ and $D_s$ mesons are plotted. The trajectories are parallel and equidistant from each other. The two-body strong decays of $D$ and $D_s$ are analyzed in the framework of heavy quark effective theory using computed masses. The strong decay widths are given in terms of strong coupling constants. These couplings are also estimated by comparing them with available experimental values for observed states. Also, the partial decay width ratios of different states are analyzed and used to suggest assignments to the observed states. We have assigned the spin-parity to newly observed $D^*_{s2}(2573)$ as the strange partner of $D^*_2(2460)$ identified as $1^3P_2$, $D_1^*(2760)$ and $D^*_{s1}(2860)$ as $1^3D_1$, $D^*_3(2750)$ and $D^*_{s3}(2860)$ as $1^3D_3$, $D_2(2740)$ as $1D_2$, $D_0(2550)$ as $2^1S_0$, $D^*_1(2660)$ and $D^*_{s1}(2700)$ as $2^3S_1$, $D^*_J(3000)$ as $2^3P_0$, $D_J(3000)$ as $2P_1$, $D^*_2(3000)$ as $1^3F_2$ states. 
\end{abstract}
\section{Introduction}
The abundance of experimental investigation of hadrons containing heavy quarks has fueled many theoretical explorations in the past decade. A multitude of new findings are anticipated in this field with the current third run of LHC, which makes a busy path for the theoretical journey of heavy hadrons. The experimentally observed $D-$ and $D_s-$ states are listed in the review of particle physics (RPP) by Particle Data Group (PDG) \cite{PDG2022}. Some of these states have confirmed $J^P$, while many others require more data to be confirmed. The $S$ and $P$ waves of non-strange and strange charm mesons are well-established states.  The $J^P$ of states $D_0(2550)$, $D^*_1(2600)$, $D_2(2740)$ and $D^*_3(2750)$ reported in the RPP \cite{PDG2022} are determined to be $0^-$, $1^-$, $2^-$ and $3^-$, respectively, by LHCb Collaboration \cite{LHCb2015,LHCb2016prd,LHCb2020}. The $D_J(3000)$ and $D_J^*(3000)$ were observed for the first time by LHCb \cite{LHCb2013} in a study of $D^+\pi^-$, $D^0\pi^+$ and $D^{*+}\pi^-$ final states. The $D_J(3000)$ meson was suggested to have unnatural parity and decaying in the $D^{*+}\pi^-$ channel final state. The other $D$ meson $D_J^*(3000)$ was reported to be decaying in the $D^+\pi^-$ and $D^0\pi^+$ final states, which suggests it to have natural parity. Later, in 2016, LHCb collaboration \cite{LHCb2016prd} also reported another new resonance $D_2^*(3000)$ in the study of $B^-\rightarrow D^+\pi^-\pi^-$ resonant substructures. The $D_2^*(3000)$ propounded to be inconsistent with the previously reported $D_J^*(3000)$, inferring they are different states. However, the LHCb collaboration also acknowledged that both resonances may have the same origin regardless of different masses and decay widths. In the strange charm ($D_s$) sector, the states $D_{sJ}(2700)^+$ and $D_{sJ}(2860)^+$ were observed by BABAR Collaboration \cite{Babar2006} in 2006. The state $D_{sJ}(2700)^+$ was confirmed by Belle Collaboration \cite{Belle2007} in 2008 with $J^P=1^-$. Also, the decays $D^*_{s1}(2700)^+\rightarrow D^* K$ and $D^*_{sJ}(2860)^+ \rightarrow D^* K$ were observed and branching fractions relative to $DK$ were measured by BABAR Collaboration \cite{babar2009} in 2009. The states  $D_{sJ}(2700)^+$ and $D_{sJ}(2860)^+$ were later reconfirmed by LHCb \cite{LHCb2012} in 2012.  The same study reported a new broad structure $D_{sJ}(3040)^+$ in the excited $D_s$ region. The LHCb Collaboration \cite{lhcb2014a,lhcb2014b} in 2014 by analyzing the $B_s^0\rightarrow \overline{D}^0K^-\pi^+$ decays observed two resonance states $D^*_{s1}(2860)^-$ with $J^P=1^-$ and $D^*_{s3}(2860)^-$ with $J^P=3^-$ in the final state of $\overline{D}^0K^-$. Therefore, the reported state $D_{sJ}(2860)$ by BABAR \cite{Babar2006,babar2009} and LHCb \cite{LHCb2012} consists of two states with $J^P$ values $1^-$ and $3^-$. In 2016, LHCb Collaboration \cite{LHCb2016jhep} reported the first observation of $D_{s2}^*(2573)^+\rightarrow D^{*+}K_s^0$ decay with a significance of $6.9\sigma$ and measured its branching fraction relative to $D^+K^0_s$ final state. Also, the presence of $D^*_{s3}(2860)^+\rightarrow D^{*+}K_s^0$ decay was reported by the LHCb. A new $D_s$ meson state $D_s(2590)^+$ is observed by LHCb Collaboration \cite{LHCb2021} in $B^0\rightarrow D^- D^+ K^+ \pi^-$ decay at a center-of-mass energy of 13 TeV. The $D_s(2590)^+$ observed decaying into the $D^+K^+\pi^-$ final state with $J^P=0^-$. The LHCb assigns this state as a strong candidate to be the $D_s(2^1S_0)^+$ state. 

Recent experimental developments have allowed studying higher radially and orbitally excited $D-$ and $D_s-$ mesons. Theoretical understanding of these recently observed mesons is an ongoing endeavor, and many phenomenological models are being used to investigate their different properties. To study masses and decays of $D-$ and $D_s-$ mesons, the heavy quark effective theory \cite{colangelo2012, wanghqet2013, Batra2015, Wanghqet2016, Pallavi2018, Pandya2021,Luo2025}, relativistic quark model \cite{DiPierro2001, songchen2015a, songchen2015b,  Licpc2022, wangprd94:2016, yuprd94:2016, godfrey2016,Yang2023}, Regge-phenomenology \cite{Liepjc2007,Chen2018}, sum rules \cite{Hayashigaki2004, Bilmis2021,Yucpc2014}, lattice QCD \cite{Moir2013lattice, kalinowski2015lattice}, effective chiral Lagrangian approach \cite{couplingdu2016}, and potential models \cite{Yazarloo2016, Alloshpot2021, akraipot2021, rashidul2023, Ruhui2022, tan2022, wang2022} are used. A detailed discussion of the theoretical understanding of $D$ and $D_s$ mesons will be given in later sections. The observed states $D^*(2300)$, $D_1(2430)$, $D_1(2420)$ and $D^*_2(2460)$ are considered to be 1P states with $J^P=0^+$, $1^+$, $1^+$ and $2^+$, respectively; however other interpretations \cite{albaladejo2017,prddu2018, PRLdu2021, Gayer2021, Asokan2023} are also possible. 

The present paper uses Song and Lin's potential \cite{linsong1987} to study the spectroscopy of the $D$ and $D_s$ mesons. The potential has previously been used to successfully study the quarkonium structures in mass spectroscopy, and their decays \cite{linsong1987, mutuk2019, Manzoor2021}. This potential is not a directly QCD-inspired potential; it contains a vector term inspired by the leptonic decay width of vector mesons and a scalar term for quark confinement at large distances.  Many other potentials are also used to study the spectroscopy and decays of heavy-light mesons \cite{rashidul2023, Yazarloo2016, akraipot2021, Alloshpot2021, Lahkar2022}. This paper extends the potential to the heavy-light system from the heavy-heavy system to which Song and Lin introduced this potential. Further, to study the strong decays of $D$ and $D_s$ mesons, we have used the heavy quark effective theory (HQET). This phenomenological effective theory combines heavy quark symmetry with chiral symmetry to analyze the strong decays of heavy hadrons through the pseudoscalar Goldstone bosons. The details are given in the further sections. This paper is organized in the following way: In section \ref{sec:framework}, we present the framework of variational method with the employed potential and the spin-dependent potential for the hyperfine splitting. The formalism of heavy quark effective theory to study the strong decay is also given in this section. The results of the calculations are discussed in the section \ref{sec:results}. The conclusions of the present analysis are given in the last section \ref{sec:conclusion}.

\section{Framework}\label{sec:framework}
\subsection{Song and Lin's Potential in variational scheme}
The Hamiltonian of a system consisting of a heavy quark of mass $m_Q$ and a light anti-quark of mass $m_{\overline{q}}$ is given below in Eq. \eqref{eq:ham} with kinetic energies of both quarks and potential energy between them. This Hamiltonian treats the kinetic energy of both heavy and light quarks relativistically.
\begin{align}
    H=\sqrt{p^2+m_Q^2}+\sqrt{p^2+m_{\overline{q}}^2}+V(r)
    \label{eq:ham}
\end{align}
The Gaussian trial wave function is used to solve for the eigen-energy of the Hamiltonian. 
\begin{align}
\label{eq:Rr}
    R_{nl}(\mu,r)=&\mu^{3/2}\left(\frac{2 \Gamma(n)}{\Gamma(n+l+1/2)}\right)^{1/2} (\mu r)^l \nonumber\\ 
    &\times e^{-\mu^2 r^2/2} L_{n-1}^{l+1/2}(\mu^2 r^2)
\end{align}
The wave function in the momentum space is the Fourier transform of Eq. \eqref{eq:Rr} given as
\begin{align}
    \label{eq:Rp}
    R_{nl}(\mu,p)=&\frac{(-1)^n}{\mu^{3/2}}\left(\frac{2 \Gamma(n)}{\Gamma(n+l+1/2)}\right)^{1/2} \left(\frac{p}{\mu}\right)^l \nonumber\\
    &\times e^{r^2/2\mu^2} L_{n-1}^{l+1/2}(p^2/\mu^2)
\end{align}
The wave functions are normalized, and a variational parameter $\mu$ is used, which will be estimated for all states.
The Song and Lin's potential \cite{linsong1987} with a constant potential depending on radial and orbital angular momentum quantum numbers is given as 
\begin{align}
    V(r)=-\frac{a}{r^{1/2}}+b r^{1/2}+V_0
    \label{eq:pot}
\end{align}
where, $V_0=a_0+cn+dl$. The $a$, $b$, $a_0$, $c$ and $d$ are potential parameters. $n$ and $l$ are the usual radial and orbital quantum numbers. In the present study, the above potential in Eq.\eqref{eq:pot} is applied on the $D-$ and $D_s-$ mesons in a variational method. The variational method we have employed to study the heavy-light mesons in the present work is inspired by Ref. \cite{Hwang1995}. Later, this method was used by Ref. \cite{akrai2002} to study the masses and decay constants of heavy-light mesons. We have extended from these studies in the sense that we used a more general wave function to study higher radial and orbital excited states. The total energy of the system is given by the expectation of the Hamiltonian $H$ as
\begin{align}
    \langle \psi | H | \psi \rangle = E(\mu)
\end{align}
The variational parameter $\mu$ is estimated by minimizing the total energy $\frac{d E(\mu)}{d\mu}=0$ at $\mu=\bar{\mu}$.  The masses taken from \cite{PDG2022} of $1S$, $1P$, and $2S$ states of $D-$mesons and $1S$ state of $D_s-$mesons are used to fix the potential parameters and masses of quarks ($m_Q$ and $m_q$). The parameters are $a=1.17$  GeV$^{\frac{1}{2}}$, $b=0.11$ GeV$^{-\frac{1}{2}}$, $a_0=-0.401$ GeV, $c=0.33$ GeV, $d=0.20$ GeV, quark masses $m_c=2.31$ GeV, $m_{u/d}=0.36$ GeV and $m_s=0.51$ GeV. The heavy quark mass $m_c$ is higher than other models \cite{Kher2017, akrai2002, akraipot2021, godfrey2016}. The present potential is not much explored for heavy-light systems, and the higher heavy quark mass may hint that this potential is better suited for bottom mesons. This will be analyzed in a future study. The obtained spin-averaged masses for $D-$ and $D_s-$ mesons and corresponding variational parameter $\mu$ are given in Table \ref{tab:CwDmass} and \ref{tab:CwDsmass} respectively. The spin-averaged masses $M_{SA,J}$ of mesons are given as
\begin{align}
    M_{SA,J}=\frac{\sum_{J}(2J+1)M_{J}}{\sum_J(2J+1)}
\end{align}
The obtained masses are in fair agreement with the other theoretical and available experimental masses in Tables \ref{tab:CwDmass} and \ref{tab:CwDsmass}. This is a good sign for the application of Song and Lin potential in heavy-light systems. We also appreciate the consistency of obtained masses by plotting Regge-trajectories, which gives good results. The variational parameter $\mu$ for all masses is also in good agreement with other models shown in Tables \ref{tab:CwDmass} and \ref{tab:CwDsmass}.
The hyperfine splitting is added perturbatively using a potential distinguishing between electric and magnetic parts of the interaction between quark and anti-quark pair \cite{cahn2003}.
\begin{align}
    V_{SD}(r)= &\frac{1}{6m_Q m_{\bar{q}}} \nabla^2 V_V (\mathbf{S_Q}.\mathbf{S_{\bar{q}}})+ \frac{1}{r}\left(\frac{dV_{V}}{dr}-\frac{dV_{S}}{dr}\right)\left(\frac{\mathbf{L.S_{Q}}}{4m_Q^2}+\frac{\mathbf{L.S_{\bar{q}}}}{4m_{\bar{q}}}\right) \nonumber\\ &+ \frac{dV_V}{dr} \left(\frac{\mathbf{L.(S_Q+S_{\bar{q}})}}{2m_Qm_{\bar{q}}} \right) + \frac{1}{12m_Q m_{\bar{q}}}\left(\frac{1}{r} \frac{dV_V}{dr}
    - \frac{d^2V_V}{dr^2} \right) \mathbf{S}_{12}
    \label{eq:sdpot}
\end{align}
The vector potential $V_V$ is the first term of the potential in Eq \eqref{eq:pot} and the scalar potential $V_S$ is the second term of potential in Eq \eqref{eq:pot}.
The 1$^{\text{st}}$ term in spin-dependent potential $V_{SD}(r)$ describes the spin-spin interaction between quark and light anti-quark pair. This term vanishes for $l\ne 0$ and can be easily calculated by taking the expectation of the Laplacian of vector potential $V_V$. The 2$^{\text{nd}}$ and 3$^{\text{rd}}$ terms consist of the interactions of orbital angular momentum ($\boldsymbol{L}$) with spin of heavy quark ($\boldsymbol{S}_Q$) and anti-quark ($\boldsymbol{S}_{\overline{q}}$). The last term describes the tensor interactions, where tensor operator $\boldsymbol{S}_{12}=2\left[3\left(\boldsymbol{S}_Q.\hat{r}\right)\left(\boldsymbol{S}_{\overline{q}}.\hat{r}\right)-\boldsymbol{S}_Q.\boldsymbol{S}_{\overline{q}}\right]$ is given by \cite{kwong1988}, 
\begin{align}
    \langle\boldsymbol{S}_{12}\rangle=&-\frac{1}{(2l-1)(2l+3)} \nonumber\\
    &\times\left[12\langle \boldsymbol{L}.\boldsymbol{S}\rangle^2+6\langle \boldsymbol{L}.\boldsymbol{S}\rangle-4\langle \boldsymbol{S}^2\rangle\langle \boldsymbol{L}^2\rangle\right]
\end{align}  

The sum of individual spins of heavy quark and light anti-quark gives the total spin momentum $\boldsymbol{S}= \boldsymbol{S}_Q + \boldsymbol{S}_{\overline{q}}$. Also, the sum of the orbital angular momentum and the spin of the light antiquark makes the light angular momentum $\boldsymbol{s}_l = \boldsymbol{L} + \boldsymbol{S}_{\overline{q}}$. The total angular momentum is the sum of light angular momentum and heavy-quark spin, $\boldsymbol{J} = \boldsymbol{s}_l + \boldsymbol{S}_Q$. 
The notations for orbital and spin angular momentum are usual $\langle \boldsymbol{L}^2\rangle=l(l+1)$, and $\langle \boldsymbol{S}^2\rangle=s(s+1)$. For $l=1$ the states with $s_l=l+s_{\overline{q}}=\frac{3}{2}$ and $J=s_l+s_Q=2$ are denoted as $^{2s+1}l_J=~^3P_2$ and the states with $s_l=l-s_{\overline{q}}=\frac{1}{2}$ and $J=s_l-s_Q=0$ are denoted as $^1P_0$. States with different total spins ($s$) and the same total angular momentum ($J$) get mixed through the spin-orbit interaction potential given in Eq. \eqref{eq:sdpot}. The physical states for P-wave ($l=1$) with $J=1$ are given by the linear combinations of the states with $J=s_l+s_Q=1$ for $s_l=l-s_{\overline{q}}=\frac{1}{2}$ ($^3P_1$) and $J=s_l-s_Q=1$ for $s_l=l+s_{\overline{q}}=\frac{3}{2}$ ($^1P_1$) as
\begin{align}
    P_1 &=~^1P_1 \text{cos}\theta_{nP} + ~^3P_1 \text{sin}\theta_{nP}\\
    P'_1 &= -~^1P_1 \text{sin}\theta_{nP}+~^3P_1 \text{cos}\theta_{nP}
\end{align}
Similar mixing expressions can be obtained for higher orbitally excited states (D and F waves). The higher mass state is denoted as $P'_1$, and the lower mass state is $P_1$. In the heavy-quark limit $m_Q\rightarrow\infty$, the light angular momentum $s_l$ and $s_Q$ are good quantum numbers. This mixing occurs due to the non-diagonal spin-orbit and tensor terms in Eq. \eqref{eq:sdpot}. The readers may go through these Ref \cite{godfrey2016, eitchenquigg1994, Gershtein_1995, cahn2003} for more clarification. The mass shift due to the spin potential terms for $P$ and $D$ waves are given in \cite{eitchenquigg1994, Gershtein_1995}. 
The expectation values are taken with the radial wavefunction $R_{nl}(r)$ given in Eqn \eqref{eq:Rr} with $\mu=\bar{\mu}$ given in Tables \ref{tab:CwDmass} and \ref{tab:CwDsmass}. 
The physical state masses for $D$ and $D_s$ mesons are enumerated in the Tables \ref{tab:SPwaveDmeson}, \ref{tab:DwaveDmeson}, \ref{tab:SPwaveDsmeson}, and \ref{tab:DwaveDsmeson} for $l=0, 1, 2$ and 3 states with their mixing angles mentioned in the caption for the corresponding states. 
The masses are in good agreement with the available experimental and other theoretical masses. 
The strong running coupling constant $\alpha_s$ can also be incorporated in the model by Ref. \cite{linsong1987}:
\begin{align}
    \alpha_s(Q^2) =  \frac{4\pi}{\beta_0~ln\left(\frac{Q^2}{\Lambda^2}\right)+\frac{\beta_1}{\beta_0}~ln\left(ln\left(\frac{Q^2}{\Lambda^2}\right)\right)}
\end{align}
where $\beta_0=11-(2/3)n_f$; $ \beta_1= 102-(38/3)n_f$; $\Lambda=100$ MeV; $Q$ is the reduced mass of the system. For charm quark $n_f=3$. The computed value of $\alpha_s (Q^2)=0.478$ using the estimated quark masses.
\begin{table}[th]
    \centering
\caption{The spin-averaged masses $(M_{SA})$ of $D$-meson states. Masses and $\bar{\mu}$ are in GeV.}
\label{tab:CwDmass}
    \begin{tabular}{cccccccc} \toprule
      State & $\bar{\mu}$ & Mass  & Ref \cite{godfrey2016} & Ref \cite{Ruhui2022} & Ref \cite{Kher2017} & Ref \cite{akraipot2021} & Exp \cite{PDG2022}  \\ \midrule
        1S & 0.4093 & 1.971 & 2.000 & 1.972 & 1.975 & 1.975 & 1.972\\
        2S & 0.2389 & 2.610 & 2.627 & 2.614 & 2.636 & 2.624 & 2.611\\
        3S & 0.1868 & 3.101 & 3.099 & 3.077 & 3.225 & 3.118 &\\
        4S & 0.1606 & 3.544 & 3.490 &       & 3.778 & 3.512 &\\
        5S & 0.1444 & 3.962 & 3.831 &       & 4.310 & 3.831 &\\
        1P & 0.2720 & 2.434 & 2.473 & 2.443 & 2.440 & 2.448 & 2.430\\
        2P & 0.2008 & 2.943 & 2.947 & 2.928 & 3.027 & 2.977 &\\
        3P & 0.1685 & 3.393 & 3.348 &       & 3.575 & 3.396 &\\
        4P & 0.1495 & 3.815 & 3.698 &       &       & 3.736 &\\
        5P & 0.1368 & 4.221 &       &       &       & 4.015 &\\
        1D & 0.2189 & 2.777 & 2.829 & 2.782 & 2.779 & 2.777 &\\
        2D & 0.1781 & 3.237 & 3.229 & 3.189 & 3.338 & 3.242 &\\
        3D & 0.1555 & 3.664 & 3.582 &       & 3.870 & 3.614 &\\
        4D & 0.1409 & 4.073 &       &       &       & 3.918 &\\
        5D & 0.1305 & 4.471 &       &       &       & 4.167 &\\ 
        1F & 0.1909 & 3.075 & 3.123 & 3.065 &       & 3.048 &\\
        2F & 0.1637 & 3.509 & 3.477 &       &       & 3.469 &\\
        3F & 0.1467 & 3.922 &       &       &       & 3.805 &\\ \bottomrule
    \end{tabular}    
\end{table}
\begin{table}[th]
    \centering
\caption{The spin-averaged masses $(M_{SA})$ of $D_s$-meson states. Masses and $\bar{\mu}$ are in GeV.}
\label{tab:CwDsmass}
    \begin{tabular}{cccccccc}\toprule
      State & $\bar{\mu}$ & Mass & Ref. \cite{godfrey2016} & Ref. \cite{Ruhui2022}  & Ref \cite{Kher2017} & Ref \cite{Yang2023} & Exp \cite{PDG2022}  \\ \midrule \midrule
        1S & 0.4453 & 2.074 & 2.091 & 2.076 & 2.076 & 2076 & 2.076\\
        2S & 0.2586 & 2.718 & 2.717 & 2.715 & 2.709 & 2906 & \\
        3S & 0.2014 & 3.211 & 3.183 & 3.175 & 3.261 &      & \\
        4S & 0.1725 & 3.653 & 3.568 &       & 3.772 &      & \\
        5S & 0.1546 & 4.072 & 3.907 &       & 4.260 &      & \\
        1P & 0.2974 & 2.542 & 2.563 & 2.542 & 2.542 & 2766 & 2.512\\
        2P & 0.2174 & 3.054 & 3.034 & 3.024 & 3.090 & 3219 & \\
        3P & 0.1815 & 3.503 & 3.429 &       & 3.595 &      &  \\
        4P & 0.1604 & 3.925 & 3.775 &       &       &      & \\
        5P & 0.1463 & 4.330 &       &       &       &      & \\
        1D & 0.2389 & 2.886 & 2.917 & 2.877 & 2.871 & 3109 & \\
        2D & 0.1925 & 3.347 & 3.310 & 3.283 & 3.383 &      & \\
        3D & 0.1672 & 3.774 & 3.660 &       & 3.869 &      & \\
        4D & 0.1509 & 4.183 &       &       &       &      & \\
        5D & 0.1394 & 4.580 &       &       &       &      & \\
        1F & 0.2080 & 3.184 & 3.199 & 3.156 &       &      & \\
        2F & 0.1769 & 3.619 & 3.552 &       &       &      & \\
        3F & 0.1577 & 4.032 &       &       &       &      & \\ \bottomrule
    \end{tabular}   
\end{table}
\begin{table}[tbh!]
    \centering
\caption{Masses of $S$ and $P$ waves of $D-$meson with hyperfine splitting. The mixing angles are: $\theta_{1P}$ = -20.32\textdegree, $\theta_{2P}$ = -20.79\textdegree, $\theta_{3P}$ = -20.87\textdegree, $\theta_{4P}$ = -20.80\textdegree and  $\theta_{5P}$ = -20.69\textdegree. All masses are in MeV. }
\label{tab:SPwaveDmeson}
    \begin{tabular*}{\textwidth}{cccccccc}\toprule
     Sates ($n^{2J+1}L_J$) & Mass & Exp \cite{PDG2022} & Ref \cite{godfrey2016} & Ref \cite{Ruhui2022}  &Ref \cite{Ebert2010} & Ref \cite{Kher2017} &Ref \cite{akraipot2021}\\ \midrule \midrule
    $1^1S_0$ & 1862 & $1864.91\pm0.05$ & 1877 & 1865 & 1871 & 1884 & 1889\\
    $1^3S_1$ & 2007 & $2008.55\pm0.03$ & 2041 & 2008 & 2010 & 2010 & 2007\\
    $2^1S_0$ & 2563 & $2549\pm19$ & 2581 & 2547 & 2581 & 2585 & 2601\\
 $2^3S_1$ & 2625 & $2627\pm10$ & 2643 & 2636 & 2632 & 2655 &2631\\
    $3^1S_0$ & 3070 &  & 3068 & 3029 & 3062 & 3186 & 3108\\
    $3^3S_1$ & 3112 &  & 3110 & 3093 & 3096 & 3239 & 3122\\
    $4^1S_0$ & 3520 & & 3468 &      & 3452 & 3746 & 3506\\
    $4^3S_1$ & 3552 & & 3497 &      & 3482 & 3789 & 3514\\
    $5^1S_0$ & 3942 & & 3814 &      & 3793 & 4283 & 3827\\
    $5^3S_1$ & 3970 & & 3837 &      & 3822 & 4319 & 3832\\
    $1^3P_0$ & 2312 & $2343\pm10$ & 2399 & 2313 & 2406 & 2357 & 2382\\
    $1P_1$   & 2442 & $2412\pm9$ & 2456 & 2424 & 2426 & 2425 & 2448\\
  $1P^{'}_1$ & 2460 & $2422.1\pm0.8$ & 2467 & 2453 & 2469 & 2447 & 2450\\
    $1^3P_2$ & 2482 & $2461.1\pm0.7$ & 2502 & 2475 & 2460 & 2461 & 2462\\
    $2^3P_0$ & 2881 & $3008.1\pm4.0$ & 2931 & 2849 & 2919 & 2976 & 2937\\
    $2P_1$   & 2947 & $2971.8\pm8.7$ & 2924 & 2900 & 2932 & 3016 & 2978\\
  $2P^{'}_1$ & 2956 & $2971.8\pm8.7$ & 2961 & 2936 & 3021 & 3034 & 2979\\
    $2^3P_2$ & 2968 & $3008.1\pm4.0$ & 2957 & 2955 & 3012 & 3039 & 2985\\
    $3^3P_0$ & 3350 & & 3343 &      & 3346 & 3536 & 3367\\
    $3P_1$   & 3396 & & 3328 &      & 3365 & 3567 & 3397\\
  $3P^{'}_1$ & 3402 & & 3360 &      & 3461 & 3582 & 3398\\
    $3^3P_2$ & 3410 & & 3353 &      & 3407 & 3584 & 3402\\
    $4^3P_0$ & 3781 & & 3697 &      &      &      & 3713\\
    $4P_1$   & 3818 & & 3681 &      &      &      & 3737\\
  $4P^{'}_1$ & 3822 & & 3709 &      &      &      & 3738\\
    $4^3P_2$ & 3830 & & 3701 &      &      &      & 3741\\
    $5^3P_0$ & 4192 & &      &      &      &      & 3996\\
    $5P_1$   & 4223 & &      &      &      &      & 4016\\
  $5P^{'}_1$ & 4227 & &      &      &      &      & 4016\\
    $5^3P_2$ & 4233 & &      &      &      &      & 4019\\ \bottomrule
    \end{tabular*}
\end{table}
\begin{table}[tbh!]
    \centering
\caption{ Masses of $D$ and $F$ wave of $D-$meson with hyperfine splitting. The mixing angles are: $\theta_{1D}$ = -68.97,$\theta_{2D}$ = -58.60, $\theta_{3D}$ = -57.53, $\theta_{4D}$ = -61.53, $\theta_{5D}$ = -66.47, $\theta_{1F}$ = -2.38, $\theta_{2F}$ = -2.11, $\theta_{3F}$ = -2.06. All masses are in MeV.  }
\label{tab:DwaveDmeson}
    \begin{tabular*}{\textwidth}{@{\extracolsep{\fill}}cccccccc} \toprule
 Sates ($n^{2J+1}L_J$) & Mass & Exp \cite{PDG2022} & Ref \cite{godfrey2016} &Ref \cite{Ruhui2022}  &Ref \cite{Ebert2010} & Ref \cite{Kher2017} &Ref \cite{akraipot2021}\\ \midrule\midrule
    $1^3D_1$ & 2759 & & 2817 & 2754 & 2788 & 2755 & 2751\\
    $1D_2$   & 2773 & & 2816 & 2755 & 2806 & 2754 & 2754\\
  $1D^{'}_2$ & 2776 & $2747\pm6$ & 2845 & 2827 & 2850 & 2783 & 2782\\
    $1^3D_3$ & 2785 & $2763.1\pm3.2$ & 2833 & 2782 & 2863 & 2788 & 2807\\
    $2^3D_1$ & 3226 & & 3231 & 3143 & 3228 & 3315 & 3223\\
    $2D_2$   & 3235 & & 3212 & 3168 & 3259 & 3318 & 3224\\
  $2D^{'}_2$ & 3236 & & 3248 & 3221 & 3307 & 3341 & 3246\\
    $2^3D_3$ & 3242 & & 3226 & 3202 & 3335 & 3355 & 3265\\
    $3^3D_1$ & 3656 & & 3588 &      &      & 3850 & 3600\\
    $3D_2$   & 3662 & & 3566 &      &      & 3854 & 3600\\
  $3D^{'}_2$ & 3664 & & 3600 &      &      & 3873 & 3617\\
    $3^3D_3$ & 3668 & & 3579 &      &      & 3885 & 3631\\
    $4^3D_1$ & 4066 & &      &      &      &      & 3906\\
    $4D_2$   & 4072 & &      &      &      &      & 3906\\
  $4D^{'}_2$ & 4073 & &      &      &      &      & 3920\\
    $4^3D_3$ & 4076 & &      &      &      &      & 3933\\
    $5^3D_1$ & 4464 & &      &      &      &      & 4158\\
    $5D_2$   & 4470 & &      &      &      &      & 4158\\
  $5D^{'}_2$ & 4470 & &      &      &      &      & 4169\\
    $5^3D_3$ & 4473 & &      &      &      &      & 4180\\
    $1^3F_2$ & 3070  & $3214\pm29\pm49$ & 3132 & 3096 & 3090 &      & 3080\\
    $1F_3$   & 3070 & & 3108 & 3022 & 3129 &      & 3051\\
  $1F^{'}_3$ & 3077 & & 3143 & 3129 & 3145 &      & 3048\\
    $1^3F_4$ & 3078 & & 3113 & 3034 & 3187 &      & 3029\\
    $2^3F_2$ & 3505 & & 3490 &      &      &      & 3494\\
    $2F_3$   & 3505 & & 3461 &      &      &      & 3472\\
  $2F^{'}_3$ & 3511 & & 3498 &      & 3551 &      & 3469\\
    $2^3F_4$ & 3511 & & 3466 &      & 3610 &      & 3454\\
    $3^3F_2$ & 3919 & &      &      &      &      & 3825\\
    $3F_3$   & 3919 & &      &      &      &      & 3807\\
  $3F^{'}_3$ & 3924 & &      &      &      &      & 3805\\
    $3^3F_4$ & 3924 & &      &      &      &      & 3793\\ \bottomrule
    \end{tabular*}  
\end{table}
\begin{table}[tbh!]
    \centering
\caption{Masses of $S$ and $P$ waves of $D_s$-meson with hyperfine splitting. The mixing angle are: $\theta_{1P}$ = -23.61, $\theta_{2P}$ = -24.00, $\theta_{3P}$ = -24.09, $\theta_{4P}$ = -24.08, $\theta_{5P}$ = -24.01. All masses are in MeV.  }
\label{tab:SPwaveDsmeson}
    \begin{tabular*}{\textwidth}{@{\extracolsep{\fill}}cccccccc} \toprule
     Sates ($n^{2J+1}L_J$) & Mass & Exp \cite{PDG2022} & Ref \cite{godfrey2016} & Ref \cite{Ruhui2022}  &Ref \cite{Ebert2010} & Ref \cite{Kher2017} & Ref \cite{Yang2023}\\ \midrule\midrule
    $1^1S_0$ & 2014 & $1969.0\pm1.4$ & 1979 & 1969 & 1969 & 1965 & 1968\\ 
    $1^3S_1$ & 2094 & $2112.2\pm0.4$ & 2129 & 2112 & 2111 & 2120 & 2112\\ 
    $2^1S_0$ & 2682 & & 2673 & 2649 & 2688 & 2680 & 2646\\ 
    $2^3S_1$ & 2730 & $2714\pm5$ & 2732 & 2737 & 2731 & 2719 & 2722\\ 
    $3^1S_0$ & 3184 & & 3154 & 3126 & 3219 & 3247 &\\ 
    $3^3S_1$ & 3220 & &3193 & 3191 & 3242 & 3265 &\\ 
    $4^1S_0$ & 3632 & & 3547 &      & 3652 & 3765 &\\ 
    $4^3S_1$ & 3661 & &3575 &      & 3669 & 3775 &\\ 
    $5^1S_0$ & 4053 & & 3894 &      & 4033 & 4280 &\\ 
    $5^3S_1$ & 4078 & & 3912 &      & 4048 & 4318 &\\ 
    $1^3P_0$ & 2440 & $2317.8\pm0.5$ & 2484 & 2409 & 2509 & 2438 & 2316\\ 
    $1P_1$   & 2544 & $2459.5$ & 2549 & 2528 & 2536 & 2529 & 2504\\ 
  $1P^{'}_1$ & 2560 & $2535.21\pm0.28$ & 2556 & 2545 & 2574 & 2541 & 2456\\ 
    $1^3P_2$ & 2581 & $2569.1\pm0.8$ & 2592 & 2575 & 2571 & 2569 & 2569\\ 
    $2^3P_0$ & 3003 &  & 3005 & 2940 & 3054 & 3022 & 2899\\ 
    $2P_1$   & 3054 & $3044\pm8^{+30}_{-5}$ & 3018 & 3002 & 3067 & 3081 & 3069\\ 
  $2P^{'}_1$ & 3062 & & 3038 & 3026 & 3154 & 3092 & 2979\\ 
    $2^3P_2$ & 3073 & & 3048 & 3053 & 3142 & 3109 & 3134\\ 
    $3^3P_0$ & 3469 & & 3439 &      & 3513 & 3541 &\\ 
    $3P_1$   & 3504 & & 3416 &      & 3519 & 3587 &\\ 
  $3P^{'}_1$ & 3510 & & 3433 &      & 3618 & 3596 &\\ 
    $3^3P_2$ & 3517 & & 3439 &      & 3580 & 3609 &\\ 
    $4^3P_0$ & 3898 & & 3764 &      &      &      &\\ 
    $4P_1$   & 3926 & & 3764 &      &      &      &\\ 
  $4P^{'}_1$ & 3930 & & 3778 &      &      &      &\\ 
    $4^3P_2$ & 3936 & & 3783 &      &      &      &\\ 
    $5^3P_0$ & 4307 & &      &      &      &      &\\ 
    $5P_1$   & 4331 & &     &      &      &      &\\ 
  $5P^{'}_1$ & 4334 & &     &      &      &      &\\ 
    $5^3P_2$ & 4340 & &     &      &      &      &\\  \bottomrule
    \end{tabular*}   
\end{table}
\begin{table}[tbh!]
    \centering
\caption{Masses of $D$ and $F$ waves of $D_s$-meson with hyperfine splitting. The mixing angles are: $\theta_{1D}$ = -78.06, $\theta_{2D}$ = -75.34, $\theta_{3D}$ = -74.57, $\theta_{4D}$ = -74.94, $\theta_{5D}$ = -75.77, $\theta_{1F}$ = -2.38, $\theta_{2F}$ = -2.11, $\theta_{3F}$ = -2.06. All masses are in MeV}
\label{tab:DwaveDsmeson}
    \begin{tabular*}{\textwidth}{@{\extracolsep{\fill}}ccccccccc} \toprule
    Sates ($n^{2J+1}L_J$) & Mass & Exp \cite{PDG2022} & Ref \cite{godfrey2016} & Ref \cite{Ruhui2022} & Ref \cite{Ebert2010}  & Ref \cite{Kher2017} & Ref \cite{Yang2023} &\\ \midrule\midrule
    $1^3D_1$ & 2868 & $2859\pm12\pm24$ & 2899 & 2843 & 2913 & 2882 & 2846 &\\
    $1D_2$   & 2881 &  & 2900 & 2857 & 2931 & 2853 & 2858 &\\
  $1D^{'}_2$ & 2886 & & 2926 & 2911 & 2961 & 2872 & 2853 &\\
    $1^3D_3$ & 2895 & $2860.5\pm2.6\pm6.5$ & 2917 & 2882 & 2971 & 2860 & 2868 &\\
    $2^3D_1$ & 3336 & & 3306 & 3233 & 3383 & 3394 & &\\ 
    $2D_2$   & 3344 & & 3298 & 3267 & 3403 & 3368 & &\\
  $2D^{'}_2$ & 3347 & & 3323 & 3306 & 3456 & 3384 & &\\
    $2^3D_3$ & 3352 & & 3311 & 3299 & 3469 & 3372 & &\\
    $3^3D_1$ & 3766 & & 3658 &      &      & 3858 & &\\
    $3D_2$   & 3772 & & 3650 &      &      & 3857 & &\\
  $3D^{'}_2$ & 3774 & & 3672 &      &      & 3869 & &\\
    $3^3D_3$ & 3778 & & 3661 &      &      & 3878 & &\\
    $4^3D_1$ & 4176 & &      &      &      &      & &\\
    $4D_2$   & 4181 & &      &      &      &      & &\\
  $4D^{'}_2$ & 4182 & &      &      &      &      & &\\
    $4^3D_3$ & 4186 & &      &      &      &      & &\\
    $5^3D_1$ & 4573 & &      &      &      &      & &\\
    $5D_2$   & 4578 &  &    &      &      &      & &\\
  $5D^{'}_2$ & 4580 &  &    &      &      &      & &\\
    $5^3D_3$ & 4582 &  &    &      &      &      & &\\
    $1^3F_2$ & 3176 & & 3208 & 3176 & 3230 &      & &\\
    $1F_3$   & 3179 & & 3186 & 3123 & 3254 &      & &\\
  $1F^{'}_3$ & 3187 & & 3218 & 3205 & 3266 &      & &\\
    $1^3F_4$ & 3189 & & 3190 & 3134 & 3300 &      & &\\
    $2^3F_2$ & 3613 & & 3562 &      &      &      & &\\
    $2F_3$   & 3615 & & 3540 &      &      &      & &\\
  $2F^{'}_3$ & 3621 & & 3569 &      & 3710 &      & &\\
    $2^3F_4$ & 3623 & & 3544 &      & 3754 &      & &\\
    $3^3F_2$ & 4027 &  &    &      &      &      & &\\
    $3F_3$   & 4029 &  &    &      &      &      & &\\
  $3F^{'}_3$ & 4033 &  &    &      &      &      & &\\
    $3^3F_4$ & 4035 &  &    &      &      &      & &\\ \bottomrule
    \end{tabular*}   
\end{table}

\subsection{Strong Decays of charm- and charm-strange mesons}
We use the heavy quark effective theory (HQET) to study the strong decays of charm ($D$) and charm-strange ($D_s$) mesons. In the heavy quark limit $m_Q\rightarrow\infty$, the heavy quark spin $s_Q$ and light angular momentum $s_l$. Thus, these angular momenta ($s_Q$, $s_l$) and the total angular momentum $J$ are separately conserved. This enables the classification of heavy-light mesons as doublets with different $s_l$. As discussed above, the total angular momentum is given as $J=s_l\pm s_Q$, and each doublet contains two states, called spin partners. The parity of the doublets is given as $P=(-1)^{l+1}$, where $l$ is the orbital angular momentum. For $l=0$ and  $s_l^P=\frac{1}{2}^-$, the doublet comprises two states with $J^P=(0^-, 1^-)$, denoted as ($P$, $P^*$). For $l=1$, we have two doublets with $s_l^P=\frac{1}{2}^+$ and $\frac{3}{2}^+$. The two doublets having $J^P_{s_l}=(0^+,1^+)_{\frac{1}{2}}$ and $J^P=(1^+,2^+)_{\frac{3}{2}}$ are denoted as ($P^*_0$, $P'_1$) and ($P_1$, $P_2^*$), respectively. For $l=2$, we have $s_l^P=\frac{3}{2}^-$ and $\frac{5}{2}^-$. The doublets with $J^P=(1^-,2^-)_{\frac{3}{2}}$ are denoted as ($P^*_1$,$P_2$) and with $J^P_{s_l}=(2^-,3^-)_{\frac{5}{2}}$ are denoted as ($P'_2$,$P_3^*$). Carrying on for $l=3$, the $s_l^P=\frac{5}{2}^+$ and $\frac{7}{2}^+$. The first doublet is denoted as ($P^*_2$,$P_3$) with $J^P_{s_l}=(2^+,3^+)_{\frac{5}{2}}$. The second doublet is denoted as ($P'_3$, $P^*_4$) with $J^P_{s_l}=(3^+,4^+)_{\frac{7}{2}}$. The classification of higher radial excitation is done in the same manner; to denote the radial excitation, we use the above notations with a tilde ($\tilde{P},\tilde{P}^*$, ...). The effective Lagrangians can be constructed by introducing effective fields in each doublet to study the strong decay dynamics of the heavy-light mesons doublets. The field $H_a$ corresponds to the $s_l=\frac{1}{2}^+$ doublet ($a=u,d,s$), $S_a$ and $T_a$ to $s_l=\frac{1}{2}^+$ and $s_l\frac{3}{2}^+$, $X_a$ and $Y_a$ to $s_l=\frac{3}{2}^-$ and $s_l=\frac{5}{2}^-$, $Z_a$ and $R_a$ to the doublets $s_l=\frac{5}{2}^+$ and $s_l=\frac{7}{2}^+$, where $a=u,d,s$ is the index for light quarks. The effective fields of the mentioned doublets are given below, 
\begin{align}
    H_{a}&=\frac{1+\slashed{v}}{2}\left[P^{*}_{a\mu}\gamma^{\mu}-P_{a}\gamma_{5} \right]\\
    S_{a}&=\frac{1+\slashed{v}}{2}\left[P'^{\mu}_{1a}\gamma_{\mu}\gamma_{5}-P^{*}_{0a} \right]\\
    T^{\mu}_{a}&=\frac{1+\slashed{v}}{2}\Bigg\{P^{*\mu\nu}_{2a}\gamma_{\nu}-P_{1a\nu}\sqrt{\frac{3}{2}}\gamma_5 \bigg[g^{\mu\nu}-\frac{1}{3}\gamma^{\nu}(\gamma^{\mu}-v^{\mu}) \bigg] \Bigg\}\\
    X^{\mu}_{a}&=\frac{1+\slashed{v}}{2}\Bigg\{P^{\mu\nu}_{2a}\gamma_{5}\gamma_{\nu}-P^{*}_{1a\nu}\sqrt{\frac{3}{2}}\bigg[g^{\mu\nu}-\frac{1}{3}\gamma^{\nu}(\gamma^{\mu}-v^{\mu}) \bigg] \Bigg\}\\
    Y^{\mu\nu}_{a}&=\frac{1+\slashed{v}}{2}\Bigg\{P^{*\mu\nu\sigma}_{3a}\gamma_{\sigma}-P^{'\alpha\beta}_{2a}\sqrt{\frac{5}{3}}\gamma_{5}\bigg[g_{\alpha}^{\mu}g_{\beta}^{\nu}-\frac{1}{5}g^{\nu}_{\beta}\gamma_{\alpha}(\gamma^{\mu}-v^{\mu})-\frac{1}{5}g^{\mu}_{\alpha}\gamma_{\beta}(\gamma^{\nu}-v^{\nu}) \bigg] \Bigg\}\\
    Z^{\mu\nu}_{a}&=\frac{1+\slashed{v}}{2}\Bigg\{P^{\mu\nu\sigma}_{3a}\gamma_{5}\gamma_{\sigma}-P^{*\alpha\beta}_{2a}\sqrt{\frac{5}{3}}\bigg[g^{\mu}_{\alpha}g^{\nu}_{\beta}-\frac{1}{5}g^{\nu}_{\beta}\gamma_{\alpha}(\gamma^{\mu}+v^{\mu})-\frac{1}{5}g_{\alpha}^{\mu}\gamma_{\beta}(\gamma^{\nu}+v^{\nu}) \bigg] \Bigg\}\\
    R^{\mu\nu\rho}_{a}&=\frac{1+\slashed{v}}{2}\Bigg\{P^{*\mu\nu\sigma}_{4a}\gamma_{5}\gamma_{\sigma}-P'^{\alpha\beta\tau}_{3a}\sqrt{\frac{7}{4}} \nonumber\ \\ 
    &~~~~~~~~~~~~~~~~\times \bigg[g^{\mu}_{\alpha}g^{\nu}_{\beta}g^{\rho}_{\tau}-\frac{1}{7}g^{\nu}_{\beta}g^{\rho}_{\tau}\gamma_{\alpha}(\gamma^{\mu}-v^{\mu})-\frac{1}{7}g^{\mu}_{\alpha}g^{\rho}_{\tau}\gamma_{\beta}(\gamma^{\nu}-v^{\nu})-\frac{1}{7}g^{\mu}_{\alpha}g^{\nu}_{\beta}\gamma_{\tau}(\gamma^{\rho}-v^{\rho}) \bigg] \Bigg\}
\end{align}    
where $v$ is the meson four-velocity, conserved in the strong interactions. The operators $P$ annihilate the corresponding mesons with four-velocity $v$ \cite{colangelo2012, wanghqet2013, wang2014}. The operators include a factor of $\sqrt{m_Q}$ and have dimension $3/2$. These effective fields of mesons interact with each other through the pseudoscalar Goldstone bosons. The octet of light pseudoscalar mesons is introduced by defining the $\xi=e^{\frac{\iota \mathcal{M}}{f_\pi}}$ and $\Sigma=\xi^2$, where the matrix $\mathcal{M}$ incorporating the $\pi$, $K$ and $\eta$ mesons fields given as
\begin{align} 
\mathcal{M}=
    \begin{pmatrix}
        \sqrt{\frac{1}{2}}\pi^0+\sqrt{\frac{1}{6}}\eta & \pi^+ & K^+ \\
        \pi^- & -\sqrt{\frac{1}{2}}\pi^0+\sqrt{\frac{1}{6}}\eta & K^0 \\
        K^- & \bar{K}^0 & -\sqrt{\frac{2}{3}}\eta
    \end{pmatrix}
\end{align}
The matrix accommodates the field of light pseudoscalar $\pi$, $K$, and $\eta$ mesons. To describe the transitions $F\rightarrow HM$, where $F = H, S, T, X, Y, Z, R$ are effective fields of heavy meson doublet and $M$ is a light pseudoscalar meson, at the leading order approximation in the light meson momentum and heavy quark mass expansion, the interaction Lagrangian terms are given as
{\allowdisplaybreaks
\begin{align}
    \mathcal{L}_{HH}&=g_{HH} \Tr \left[ \bar{H}_a H_b \gamma_{\mu}\gamma_5 \mathcal{A}^{\mu}_{ba} \right]\\
    \mathcal{L}_{SH}&=g_{SH} \Tr \big[\bar{H}_a S_b \gamma_{\mu}\gamma_5 \mathcal{A}^{\mu}_{ba} \big]+\text{H.c.}\\
    \mathcal{L}_{TH}&=\frac{g_{TH}}{\Lambda_{\chi}}\Tr \big[\bar{H}_aT^{\mu}_b(\dot{\iota}D_{\mu}\slashed{\mathcal{A}}+\dot{\iota}\slashed{D}\mathcal{A}_{\mu})_{ba}\gamma_5 \big]+\text{H.c.}\\
    \mathcal{L}_{XH}&=\frac{g_{XH}}{\Lambda_{\chi}}\Tr\big[\bar{H}_a X^{\mu}_b\big(\dot{\iota}D_{\mu}\slashed{\mathcal{A}}+
    \dot{\iota}\slashed{D}\mathcal{A}_{\mu}\big)_{ba}\gamma_5 \big]+\text{H.c.}\\
    \mathcal{L}_{YH}&=\frac{1}{\Lambda^2_{\chi}}\Tr \big[\bar{H}_aY^{\mu\nu}_b\big[k^{Y}_1\{D_{\mu},D_{\nu}\}\mathcal{A}_{\lambda}+k^{Y}_2(D_{\mu}D_{\lambda}\mathcal{A}_{\nu}+D_{\nu}D_{\lambda}\mathcal{A}_{\mu})\big]_{ba}\gamma^{\lambda}\gamma_5\big]+\text{H.c.}\\ 
    \mathcal{L}_{ZH}&=\frac{1}{\Lambda_{\chi}}\Tr[\bar{H}_a Z^{\mu\nu}_b[k^Z_1\{D_{\mu},D_{\nu}\}\mathcal{A}_{\lambda}+k^Z_2(D_{\mu}D_{\lambda}A_{\nu}+D_{\nu}D_{\lambda}\mathcal{A}_{mu})]_{ba}\gamma^{\lambda}\gamma_5]+\text{H.c.}\\
    \mathcal{L}_{RH}&=\frac{1}{\Lambda^3_{\chi}}\Tr[\bar{H}_a R^{\mu\nu\rho}[k^R_1\{D_{\mu},D_{\nu},D_{\rho}\}\mathcal{A}_{\lambda}+k^R_2({D_{\mu},D_{\rho}}D_{\lambda}\mathcal{A}_{\nu}{D_{\nu},D_{\rho}}D_{\lambda}\mathcal{A}_{\mu} \nonumber\\ 
    &~~~~~~~~~~~~~~~~~~~~~~~~+{D_{\mu},D_{\nu}}D_{\lambda}\mathcal{A}_{\rho})]_{ba}\gamma^{\lambda}\gamma_5]+\text{H.c.}
\end{align}}
The definitions of identities in the above equations are $D_{\mu}=\partial_{\mu}+\mathcal{V}_{\mu}$; $\{D_{\mu},D_{\nu}\}=D_{\mu}D_{\nu}+D_{\nu}D_{\mu}$; $\{D_{\mu},D_{\nu},D_{\rho}\}=D_{\mu}D_{\nu}D_{\rho}+D_{\mu}D_{\rho}D_{\nu}+D_{\nu}D_{\mu}D_{\rho}+D_{\nu}D_{\rho}D_{\mu}+D_{\rho}D_{\mu}D_{\nu}+D_{\rho}D_{\nu}D_{\mu}$. The axial-vector ($\mathcal{A}$) and vector ($\mathcal{V}$) currents are defined as 
\begin{align}
    \mathcal{A}_{\mu}=\frac{1}{2}(\xi^{\dagger}\partial_{\mu}\xi-\xi\partial_{\mu}\xi^{\dagger})\\
    \mathcal{V}_{\mu}=\frac{1}{2}(\xi^{\dagger}\partial_{\mu}\xi+ \xi\partial_{\mu}\xi^{\dagger})
\end{align}
The chiral symmetry-breaking scale $\Lambda_{\chi}$ is set to 1 GeV. The strong coupling constants involved in their corresponding transitions are $g_{HH}$, $g_{SH}$, $g_{TH}$, $g_{XH}$, $g_{YH}=k^Y_1+k^Y_2$, $g_{ZH}=k^Z_1+k^Z_2$, $g_{RH}=k^R_1+k^R_2$. These couplings can be estimated by the experimental decay widths. The $f_{\pi}$ is taken to be 132 MeV. The decay widths from the above effective fields and the interaction Lagrangians are given below \cite{colangelo2012, Pandya2021, wang2014}. The strong decay width for two-body decay of heavy-light ($D$ or $D_s$) mesons by emitting pseudoscalar light mesons ($\pi$, $\eta$, and $K$) according to the Lagrangians $\mathcal{L}_{HH}$, $\mathcal{L}_{SH}$, $\mathcal{L}_{TH}$, $\mathcal{L}_{XH}$, $\mathcal{L}_{YH}$, $\mathcal{L}_{ZH}$ and $\mathcal{L}_{RH}$ is given as 
\begin{align}
    \Gamma&=\frac{1}{2J+1}\sum\frac{p_M}{8\pi M_i^2}|\mathcal{A}|^2 \\
    p_M&= \frac{\sqrt{(M_i^2-(M_f+M_M)^2)(M_i^2-(M_f-M_M)^2)}}{2M_i} \nonumber
\end{align}
where, $\mathcal{A}$ is the amplitude of the transition, $J$ is total angular momentum of initial heavy meson, $\sum$ represents the sum over all polarization vectors, $M_i$ is the mass of initial heavy meson, $M_f$ represents the mass of final heavy meson, $M_M$ and $p_M$ are mass and momentum of emitted light pseudoscalar meson. The explicit expressions for decay widths of transition of heavy-light mesons from different fields ($H$, $S$, $T$, $X$, $Y$, $Z$, and $R$) to $H-$field for different channels are given below \cite{colangelo2012,wang2014}.

Decays of $H$-field $(0^+,1^+)$ of $S$-wave
\begin{align}
    \Gamma(1^{-}\rightarrow 0^{-})= C_M \frac{g_{HH}^2}{6\pi f_{\pi}^{2}}\frac{M_{f}}{M_{i}}\lvert\vec{p}_M\rvert^3\\    
    \Gamma(1^{-}\rightarrow 1^{-})= C_M \frac{g_{HH}^2}{3\pi f_{\pi}^2}\frac{M_{f}}{M_{i}}\lvert\vec{p}_M\rvert^3 \\
    \Gamma(0^{-}\rightarrow 1^{-})=C_M \frac{g_{HH}^2}{2\pi f_{\pi}^2}\frac{M_{f}}{M_{i}}\lvert\vec{p}_M\rvert^3
\end{align}   

Decays of $S$-field $(0^+,1^+)$ of $P$-wave
\begin{align}
    \Gamma(0^+\rightarrow0^-)=C_M \frac{ g_{SH}^2}{2\pi f_{\pi}^2}\frac{M_f}{M_i}\left[M^{2}_{M}+\lvert\vec{p}_{M}\rvert^{2}\right]\lvert\vec{p}_M\rvert\\
    \Gamma(1^{+}\rightarrow 1^{-})=C_M \frac{ g_{SH}^2}{2\pi f_{\pi}^2}\frac{M_f}{M_i}\left[M^{2}_{M}+\lvert\vec{p}_{M}\rvert^{2}\right]\lvert\vec{p}_M\rvert
\end{align}

Decays of $T$-field $(1^+,2^+)$ of $P$-wave
\begin{align}
    \Gamma(1^{+}\rightarrow 1^{-})=C_{M}\frac{2g_{TH}^2}{3\pi f_{\pi}^2 \Lambda_{\chi}^2}\frac{M_f}{M_i}\lvert\vec{p}_M\rvert^5\\
    \Gamma(2^{+}\rightarrow 0^{-})=C_{M}\frac{4g_{TH}^2}{15\pi f_{\pi}^{2} \Lambda_{\chi}^2} \frac{M_f}{M_i}\lvert\vec{p}_M\rvert^{5}\\
    \Gamma(2^{+}\rightarrow 1^{-})=C_{M}\frac{2g_{TH}^2}{5\pi f_{\pi}^{2} \Lambda_{\chi}^2}\frac{M_f}{M_i}\lvert\vec{p}_M\rvert^{5}
\end{align}

Decays of $X$-field $(1^{-},2^{-})$ of $D$-wave
\begin{align}
    \Gamma(1^{-}\rightarrow 0^{-})=C_{M}\frac{4g_{XH}^2}{9\pi f_{\pi}^2 \Lambda_{\chi}^2}\frac{M_f}{M_i}\left[M_{M}^{2}+\lvert\vec{p}_{M}\rvert^{2}\right]\lvert\vec{p}_M\rvert^3\\
    \Gamma(1^{-}\rightarrow 1^{-})=C_{M}\frac{2g_{XH}^2}{9\pi f_{\pi}^2 \Lambda_{\chi}^2}\frac{M_f}{M_i}\left[M_{M}^{2}+\lvert\vec{p}_{M}\rvert^{2}\right]\lvert\vec{p}_M\rvert^3\\
    \Gamma(2^{-}\rightarrow 1^{-})=C_{M}\frac{2g_{XH}^2}{3\pi f_{\pi}^2 \Lambda_{\chi}^2}\frac{M_f}{M_i}\left[M_{M}^{2}+\lvert\vec{p}_{M}\rvert^{2}\right]\lvert\vec{p}_M\rvert^3
\end{align}

Decays of $Y$-field $(2^{-},3^{-})$ of $D$-wave
\begin{align}
    \Gamma(2^{-}\rightarrow 1^{-})=C_{M}\frac{4g_{YH}^2}{15\pi f_{\pi}^2 \Lambda_{\chi}^4}\frac{M_f}{M_i}\lvert\vec{p}_M\rvert^7\\
    \Gamma(3^{-}\rightarrow 0^{-})=C_{M}\frac{4g_{YH}^2}{35\pi f_{\pi}^2 \Lambda_{\chi}^4}\frac{M_f}{M_i}\lvert\vec{p}_M\rvert^7\\
    \Gamma(3^{-}\rightarrow 1^{-})=C_{M}\frac{16g_{YH}^2}{105\pi f_{\pi}^2 \Lambda_{\chi}^4}\frac{M_f}{M_i}\lvert\vec{p}_M\rvert^7
\end{align}

Decays of $Z$-field $(2^{+},3^{+})$ of $F$-wave
\begin{align}
    \Gamma(2^{+}\rightarrow 0^{-})=C_{M}\frac{4g_{ZH}^2}{25\pi f_{\pi}^2 \Lambda_{\chi}^4}\frac{M_f}{M_i}\left[M_{M}^{2}+\lvert\vec{p}_{M}\rvert^{2}\right]\lvert\vec{p}_M\rvert^5\\
    \Gamma(2^{+}\rightarrow 1^{-})=C_{M}\frac{8g_{ZH}^2}{75\pi f_{\pi}^2 \Lambda_{\chi}^4}\frac{M_f}{M_i}\left[M_{M}^{2}+\lvert\vec{p}_{M}\rvert^{2}\right]\lvert\vec{p}_M\rvert^5\\
    \Gamma(3^{+}\rightarrow 1^{-})=C_{M}\frac{4g_{ZH}^2}{15\pi f_{\pi}^2 \Lambda_{\chi}^4}\frac{M_f}{M_i}\left[M_{M}^{2}+\lvert\vec{p}_{M}\rvert^{2}\right]\lvert\vec{p}_M\rvert^5
\end{align}

Decays of $R$-field $(3^{+},4^{+})$ of $F$-wave
\begin{align}
    \Gamma(3^{+}\rightarrow 1^{-})=C_{M}\frac{36g_{RH}^2}{35\pi f_{\pi}^2 \Lambda_{\chi}^6}\frac{M_f}{M_i}\lvert\vec{p}_M\rvert^9\\
    \Gamma(4^{+}\rightarrow 0^{-})=C_{M}\frac{16g_{RH}^2}{35\pi f_{\pi}^2 \Lambda_{\chi}^6}\frac{M_f}{M_i}\lvert\vec{p}_M\rvert^9\\
    \Gamma(4^{+}\rightarrow 1^{-})=C_{M}\frac{4g_{RH}^2}{7\pi f_{\pi}^2 \Lambda_{\chi}^6}\frac{M_f}{M_i}\lvert\vec{p}_M\rvert^9\\
\end{align}
where $C_M$ is the factor dependent on the light pseudoscalar mesons, $C_{\pi_+}=C_{K^+}=1$, $C_{\pi^0}=C_{K_s}=\frac{1}{2}$, $C_{\eta}=\frac{1}{6}$ or $\frac{2}{3}$ (for initial meson $c\bar{q}$ or $c\bar{s}$ respectively). The decay widths of different candidate states for the experimentally observed $D-$ and $D_s-$ meson family are calculated and shown in Table \ref{tab:1p1d} for $1P$ and $1D$ states, Table \ref{tab:2s2ppwidth} for $2S$ and $2P$ states, Table \ref{tab:3s2dwidth} for $3S$ and $2D$ states, and Table \ref{tab:1fwidth} for $1F$ states. In the next section, we will discuss the results from this section, and some meaningful conclusions will be drawn later.
\begin{table}[tbh!]
    \centering
        \caption{Strong decay widths of $1P$ and $1D$ states of $D$ and $D_s$ mesons using the masses calculated in the present study. The ratio in 4th column represents the $\hat{\Gamma}=\frac{\Gamma}{\Gamma(D_J\rightarrow D^{*+}\pi^-)}$ for $D$ mesons and the ratio in 8th column represents the $\hat{\Gamma}=\frac{\Gamma}{\Gamma(D_{sJ}\rightarrow D^{*+}K^0)}$ for $D_s$ mesons. Fractions in the 5th and 9th columns are the percentage of the partial width to the total decay width.}
    \begin{tabular}{@{\extracolsep{\fill}}p{1cm}p{1.2cm}cccp{1.2cm}ccc}\toprule
States ($nL_{s_l}J^P$)  & Decay Mode  & Width & Ratio  & Fraction & Decay Mode & Width & Ratio & Fraction \\ \midrule \midrule
\multirow{9}{*}{$1P_{\frac{3}{2}}2^+$}   & $D^{*+}\pi^-$ & 66.59 $g_{TH}^2$    &  1    & 20.84 & 
                        $D^{*+}K^0$     & 5.25 $g_{TH}^2$ & 1       & 4.04  \\
                        & $D^{*0}\pi^0$ & 34.97 $g_{TH}^2$   & 0.52 & 10.95 & 
                        $D^{*0}K^+$ & 6.62 $g_{TH}^2$ & 1.26 & 5.10 \\
                        & $D^{*0}\eta $ &      -               & -       &  -        & 
                        $D^{*+}_{s}\eta $      & -                & -       & - \\
                        & $D^{*+}_s K^-$& -                    & -       & -         & 
                        $D^{*+}_{s}\pi^0$    & 33.80 $g_{TH}^2\times 10^{-4}$  & - & - \\
                        & $D^{+}\pi^-$  & 142.78 $g_{TH}^2$ & 2.14 & 44.69 &  
                        $D^{+}K^0$ & 55.30 $g_{TH}^2$ & 10.53 & 42.56 \\
                        & $D^{0}\pi^0$  & 74.45 $g_{TH}^2$   & 1.12 & 23.30 &  
                        $D^{0}K^+$ & 60.78 $g_{TH}^2$ & 11.58 & 46.78\\
                        & $D^{0}\eta $  & 0.56 $g_{TH}^2$     & 0.01 & 0.17  & 
                        $D^{+}_{s}\eta $       & 1.97 $g_{TH}^2$   & 0.37    & 1.51 \\
                        & $D^{+}_s K^-$ & 0.12 $g_{TH}^2$      & 0.00 & 0.04 & 
                        $D^{+}_{s}\pi^0$     & 75.11 $\times 10^{-4}$ $g_{TH}^2$ & - & - \\
                        & Total         & 319.48 $g_{TH}^2$  &         &          & 
                        Total       & 129.93 $g_{TH}^2$ &       & \\ \midrule
\multirow{9}{*}{$1D_{\frac{3}{2}}1^-$}   & $D^{*+}\pi^-$     & 321.45 $g_{XH}^2$     & 1     & 10.75 & $D^{*+}K^0$         & 350.79 $g_{XH}^2$     & 1     & 8.30  \\
                        & $D^{*0}\pi^0$     & 164.09 $g_{XH}^2$     & 0.51  & 5.49 & 
                        $D^{*0}K^+$         & 361.66 $g_{XH}^2$     & 1.03  & 8.55 \\
                        & $D^{*0}\eta $     & 21.08  $g_{XH}^2$     & 0.06  & 0.70 & 
                        $D^{*+}_{s}\eta $   & 90.76 $g_{XH}^2 $    & 0.26   & 2.15  \\
                        & $D^{*+}_s K^-$    & 58.14 $g_{XH}^2$      & 0.18  & 1.94 & 
                        $D^{*+}_{s}\pi^0$   & 0.02 $g_{TH}^2$       & - & -     \\
                        & $D^{+}\pi^-$      & 1234.53 $g_{XH}^2$    & 3.84  & 41.30 & 
                        $D^{+}K^0$          & 1560.59 $g_{XH}^2$    & 4.08  & 33.86 \\
                        & $D^{0}\pi^0$      & 630.67 $g_{XH}^2$     & 1.96  &  21.10 & 
                        $D^{0}K^+$          & 1601.90 $g_{XH}^2$    & 4.19  & 34.79 \\
                        & $D^{0}\eta $      & 120.09 $g_{XH}^2$     & 0.37  & 4.02 & 
                        $D^{+}_{s}\eta $    & 583.20 $g_{XH}^2$     & 1.49  & 12.34  \\
                        & $D^{+}_s K^-$     & 439.37 $g_{XH}^2$     & 1.37  & 14.70 & 
                        $D^{+}_{s}\pi^0$    & 0.07 $g_{XH}^2$       & -     & -  \\
                        & Total             &  2989.42 $g_{XH}^2$   &       &           
                        & Total             & 4228.3 $g_{XH}^2$    &       &  \\ \midrule 
\multirow{5}*{$1D_{\frac{3}{2}}2^-$}   & $D^{*+}\pi^-$     & 1043.28 $g_{XH}^2$    & 1     & 56.23 & 
                        $D^{*+}K^0$         & 1142.16 $g_{XH}^2$    & 1     & 43.49   \\
                        & $D^{*0}\pi^0$     & 532.27 $g_{XH}^2$     & 0.51  & 28.69 & 
                        $D^{*0}K^+$         & 1176.28 $g_{XH}^2$    & 1.03  & 44.78 \\
                        & $D^{*0}\eta $     & 72.29  $g_{XH}^2$     & 0.07  & 3.90 & 
                        $D^{*0}_{s}\eta $   & 307.96 $g_{XH}^2$     & 0.27  & 11.72  \\
                        & $D^{*+}_s K^-$    & 207.35 $g_{XH}^2$     & 0.20  & 11.18 &    
                        $D^{*+}_{s}\pi^0$   & 0.06 $g_{TH}^2$       & -     & -\\
                        & Total             &  1855.19 $g_{XH}^2$ &         & & Total & 2626.46 $g_{XH}^2$ & &  \\ \cmidrule{1-9} 
\multirow{5}{*}{$1D_{\frac{5}{2}}2^-$}   & $D^{*+}\pi^-$     &  170.71 $g_{YH}^2$    & 1      & 64.04  & 
                        $D^{*+}K^0$         & 104.37 $g_{YH}^2$     & 1     & 45.92   \\
                        & $D^{*0}\pi^0$     &  88.25 $g_{YH}^2$     & 0.52  & 33.10  & 
                        $D^{*0}K^+$         & 110.82 $g_{YH}^2$     & 1.06  & 48.75 \\
                        & $D^{*0}\eta $     & 2.64  $g_{YH}^2$      & 0.01  & 1.00   & 
                        $D^{*0}_{s}\eta $   & 12.10 $g_{YH}^2$      & 0.11  & 5.32 \\
                        & $D^{*+}_s K^-$    & 4.97  $g_{YH}^2$      & 0.03  & 1.86   & 
                        $D^{*+}_{s}\pi^0$   & 0.01 $g_{YH}^2$      & - & - \\
                        & Total             & 266.56 $g_{YH}^2$     &       &           
                        & Total             & 227.31 $g_{YH}^2$     &       &  \\ \midrule 
\multirow{9}{*}{$1D_{\frac{5}{2}}3^-$}   & $D^{*+}\pi^-$     & 104.81 $g_{YH}^2$     & 1     & 21.22 & 
                        $D^{*+}K^0$         & 65.47 $g_{YH}^2$      & 1     & 13.24   \\
                        & $D^{*0}\pi^0$     & 54.15 $g_{YH}^2$      & 0.52  & 10.96 & 
                        $D^{*0}K^+$         & 69.41 $g_{YH}^2$      & 1.06  & 14.04 \\
                        & $D^{*0}\eta $     & 1.75  $g_{YH}^2$      & 0.17 & 0.35   & 
                        $D^{*0}_{s}\eta $   & 8.02 $g_{YH}^2$       & 0.12  & 1.62 \\
                        & $D^{*+}_s K^-$    & 3.45 $g_{YH}^2$       & 0.03 & 0.70   & 
                        $D^{*+}_{s}\pi^0$   & 59.38 $\times 10^{-4}$ $g_{TH}^2$  & - & - \\
                        & $D^{+}\pi^-$      & 197.16 $g_{YH}^2$     & 1.88 & 39.92 & 
                        $D^{+}K^0$          & 154.33 $g_{TH}^2$     & 2.36   & 31.22 \\
                        & $D^{0}\pi^0$      & 101.92 $g_{YH}^2$   & 0.97 & 20.63 & 
                        $D^{0}K^+$          & 162.32 $g_{TH}^2$     & 2.48 & 32.83 \\
                        & $D^{0}\eta $      & 7.57 $g_{YH}^2$        & 0.07 & 1.53    & 
                        $D^{0}_{s}\eta $    & 34.76 $g_{TH}^2$    &  0.53    & 7.03  \\
                        & $D^{+}_s K^-$     & 23.09 $g_{YH}^2$      & 0.22 & 4.67    & 
                        $D^{+}_{s}\pi^0$    & 113.64 $\times 10^{-4}$ $g_{TH}^2$ & - & -  \\
                        & Total             & 493.89 $g_{YH}^2$   &          &            & 
                        Total               & 494.33 $g_{TH}^2$ &         & \\ \bottomrule 
    \end{tabular}
    \label{tab:1p1d}
\end{table}
\begin{table}[tbh!]
    \centering
    \caption{Strong decay widths of $2S$ and $2P$ states of $D$ and $D_s$ mesons using the masses calculated in the present study. In the 4th column, the ratio for $D$ meson states without $D^{*+}\pi^-$ channel is $\hat{\Gamma}=\frac{\Gamma}{\Gamma(D_J\rightarrow D^+\pi^-)}$. For $D_s$ meson states with $D^{*+}K^0$ channel absent, the ratio in 8th column is $\hat{\Gamma}=\frac{\Gamma}{\Gamma(D_{sJ}\rightarrow D^+K^0)}$. }
    \begin{tabular*}{1.\textwidth}{@{\extracolsep{\fill}}p{1cm}p{1.2cm}ccccccc}\toprule
States ($nL_{s_l}J^P$) & Decay Mode  & Width                & Ratio & Fraction & Decay Mode& Width                & Ratio & Fraction \\ \midrule \midrule
$2S_{\frac{1}{2}}0^-$   & $D^{*+}\pi^-$    & 777.24 $\tilde{g}_{HH}^2$  & 1     & 66.13 & 
                        $D^{*+}K^0$        & 414.00 $\tilde{g}_{HH}^2$  & 1     & 47.83     \\
                     & $D^{*0}\pi^0$       & 397.40 $\tilde{g}_{HH}^2$  & 0.51  & 33.81 & 
                     $D^{*0}K^+$           & 438.56  $\tilde{g}_{HH}^2$ & 1.06  & 50.67     \\
                     & $D^{*0}\eta $       & 0.72  $\tilde{g}_{HH}^2$   & 0.00  & 0.06  & 
                     $D^{*+}_s\eta$        & 12.89  $\tilde{g}_{HH}^2$  & 0.03  & 1.49      \\
                     & $D^{*+}_s K^-$      & -                          & -     &  -    & 
                     $D^{*+}_s\pi^0$       & 0.04 $\tilde{g}_{HH}^2$    & -     & -         \\
                     & Total               & 1175.36 $\tilde{g}_{HH}^2$ &       &       & 
                     Total                 & 865.50  $\tilde{g}_{HH}^2$ &       &           \\ \midrule
$2S_{\frac{1}{2}}1^-$   & $D^{*+}\pi^-$    & 687.79 $\tilde{g}_{HH}^2$  & 1     & 33.50 & 
                        $D^{*+}K^0$        & 405.80 $\tilde{g}_{HH}^2$  & 1     & 21.63     \\
                     & $D^{*0}\pi^0$       & 350.44 $\tilde{g}_{HH}^2$  & 0.51  & 17.07 & 
                     $D^{*0}K^+$           & 424.27 $\tilde{g}_{HH}^2$  & 1.04  & 22.62     \\
                     & $D^{*0}\eta $       & 12.24  $\tilde{g}_{HH}^2$  & 0.02  & 0.60  & 
                     $D^{*+}_s\eta$        & 50.02 $\tilde{g}_{HH}^2$   & 0.49  & 11.99     \\
                     & $D^{*+}_s K^-$      & 9.45  $\tilde{g}_{HH}^2$   & 0.12  & 2.67  & 
                     $D^{*+}_s\pi^0$       & 0.03 $\tilde{g}_{HH}^2$    & -     & -         \\
                     & $D^{+}\pi^-$        & 555.86 $\tilde{g}_{HH}^2$  & 0.81  & 27.08 & 
                     $D^{+}K^0$            & 423.43 $\tilde{g}_{HH}^2$  & 1.04  & 22.57     \\
                     & $D^{0}\pi^0$        & 282.81 $\tilde{g}_{HH}^2$  & 0.41  & 13.78 & 
                     $D^{0}K^+$            & 436.87 $\tilde{g}_{HH}^2$  & 1.08  & 23.29     \\
                     & $D^{0}\eta $        & 32.26 $\tilde{g}_{HH}^2$   & 0.05  & 1.57  & 
                     $D^{+}_s\eta$         & 135.33 $\tilde{g}_{HH}^2$  & 0.33  & 7.21      \\
                     & $D^{+}_s K^-$       & 121.97 $\tilde{g}_{HH}^2$  & 0.18  & 5.94  & 
                     $D^{+}_s\pi^0$        & 0.03 $\tilde{g}_{HH}^2$    & -     & -         \\
                     & Total               & 2052.83 $\tilde{g}_{HH}^2$ &       &       & 
                     Total                 & 1875.79 $\tilde{g}_{HH}^2$ &       &           \\ \midrule
                     
$2P_{\frac{1}{2}}0^+$   & $D^{+}\pi^-$     & 3429.86 $\tilde{g}_{SH}^2$ & 1     & 41.13 & 
                        $D^{+}K^0$         & 4313.41 $\tilde{g}_{SH}^2$ & 1     & 39.43     \\
                        & $D^{0}\pi^0$     & 1730.14 $\tilde{g}_{SH}^2$ & 0.50  & 20.75 & 
                        $D^{0}K^+$         & 4350.20 $\tilde{g}_{SH}^2$ & 1.01  & 39.43     \\
                        &$D^{0}\eta $      & 545.35 $\tilde{g}_{SH}^2$  & 0.16  & 6.54  & 
                        $D^{+}_s\eta$      & 2367.74 $\tilde{g}_{SH}^2$ & 0.55  & 21.46     \\
                        & $D^{+}_s K^-$    & 2633.93 $\tilde{g}_{SH}^2$ & 0.77  & 31.58 & 
                        $D^{+}_s\pi^0$     & 0.19  $\tilde{g}_{SH}^2$   & -     &-          \\
                        & Total            & 8339.29 $\tilde{g}_{SH}^2$ &       &       & 
                        Total              & 11031.50 $\tilde{g}_{SH}^2$&       &           \\ \midrule
$2P_{\frac{1}{2}}1^+$   & $D^{*+}\pi^-$    & 3037.28 $\tilde{g}_{SH}^2$ & 1     & 42.20 & 
                        $D^{*+}K^0$        & 3735.50 $\tilde{g}_{SH}^2$ & 1     & 39.57     \\
                        & $D^{*0}\pi^0$    & 1529.94 $\tilde{g}_{SH}^2$ & 0.50  & 21.25 & 
                        $D^{*0}K^+$        & 3764.54 $\tilde{g}_{SH}^2$ & 1.01  & 39.87     \\
                        & $D^{*0}\eta $    & 468.09 $\tilde{g}_{SH}^2$  & 0.15  & 6.50  & 
                        $D^{*+}_s\eta$     & 1940.79 $\tilde{g}_{SH}^2$ & 0.52  & 20.56     \\
                        & $D^{*+}_s K^-$   & 2162.68 $\tilde{g}_{SH}^2$ & 0.71  & 30.04 & 
                        $D^{*+}_s\pi^0$    & 0.16 $\tilde{g}_{SH}^2$    & -     & -         \\
                        & Total            & 7197.99 $\tilde{g}_{SH}^2$ &       &       & 
                        Total              & 9440.99 $\tilde{g}_{SH}^2$ &       &           \\ \midrule
$2P_{\frac{3}{2}}1^+$   & $D^{*+}\pi^-$    & 2475.87 $\tilde{g}_{TH}^2$ & 1     & 55.25 & 
                        $D^{*+}K^0$        & 2077.75 $\tilde{g}_{TH}^2$ & 1     & 42.64     \\
                        & $D^{*0}\pi^0$    & 1259.09 $\tilde{g}_{TH}^2$ & 0.51  & 28.10 & 
                        $D^{*0}K^+$        & 2135.35 $\tilde{g}_{TH}^2$ & 1.03  & 43.82     \\
                        & $D^{*0}\eta $    & 155.83 $\tilde{g}_{TH}^2$  & 0.06  & 3.48  & 
                        $D^{*+}_s\eta$     & 659.15 $\tilde{g}_{TH}^2$  & 0.32  & 13.53     \\
                        & $D^{*+}_s K^-$   & 590.40 $\tilde{g}_{TH}^2$  & 0.23  & 13.17 & 
                        $D^{*+}_s\pi^0$    & 0.13   $\tilde{g}_{TH}^2$  & -     &   -       \\
                        & Total            & 4481.19 $\tilde{g}_{TH}^2$ &       &       & 
                        Total              & 4872.38 $\tilde{g}_{TH}^2$ &       &           \\ \midrule 
$2P_{\frac{3}{2}}2^+$   & $D^{*+}\pi^-$    & 1565.12 $\tilde{g}_{TH}^2$ & 1     & 25.39 & 
                        $D^{*+}K^0$        & 1318.78 $\tilde{g}_{TH}^2$ & 1     & 19.35     \\
                        & $D^{*0}\pi^0$    & 795.68 $\tilde{g}_{TH}^2$  & 0.51  & 12.91 & 
                        $D^{*0}K^+$        & 1354.50 $\tilde{g}_{TH}^2$ & 1.03  & 19.87     \\
                        & $D^{*0}\eta $    & 101.44 $\tilde{g}_{TH}^2$  & 0.06  & 1.64  & 
                        $D^{*+}_s\eta$     & 426.30 $\tilde{g}_{TH}^2$  & 0.32  & 6.25      \\
                        & $D^{*+}_s K^-$   & 389.24 $\tilde{g}_{TH}^2$  & 0.25  & 6.31  & 
                        $D^{*+}_s\pi^0$    & 0.08 $\tilde{g}_{TH}^2$    & -     & -         \\
                        & $D^{+}\pi^-$     & 1689.59 $\tilde{g}_{TH}^2$ & 1.08  & 27.41 & 
                        $D^{+}K^0$         & 1534.53 $\tilde{g}_{TH}^2$ & 1.16  & 22.51     \\
                        & $D^{0}\pi^0$     & 859.65 $\tilde{g}_{TH}^2$  & 0.55  & 13.94 & 
                        $D^{0}K^+$         & 1573.39 $\tilde{g}_{TH}^2$ & 1.19  & 23.08     \\
                        & $D^{0}\eta $     & 142.20 $\tilde{g}_{TH}^2$  & 0.91  & 2.31  & 
                        $D^{+}_s\eta$      & 608.95 $\tilde{g}_{TH}^2$  & 0.46  & 8.93      \\
                        & $D^{+}_s K^-$    & 621.81 $\tilde{g}_{TH}^2$  & 0.40  & 10.09 & 
                        $D^{+}_s\pi^0$     & 0.09 $\tilde{g}_{TH}^2$    & -     & -         \\
                        & Total            & 6164.73 $\tilde{g}_{TH}^2$ &       &       & 
                        Total              & 6816.63 $\tilde{g}_{TH}^2$ &       &           \\ \bottomrule
    \end{tabular*}
    \label{tab:2s2ppwidth}
\end{table}

\begin{table}[tbh!]
    \centering
    \caption{Strong decay widths of $3S$ and $2D$ states of $D$ and $D_s$ mesons using the masses calculated in the present study.}
    \begin{tabular*}{\textwidth}{@{\extracolsep{\fill}}p{1cm}p{1.2cm}cccp{1.2cm}ccc}\toprule
    States ($nL_{s_l}J^P$) & Decay Mode  & Width   & Ratio & Fraction & Decay Mode& Width                & Ratio & Fraction \\ \midrule \midrule
$3S_{\frac{1}{2}}0^-$   & $D^{*+}\pi^-$  & 3923.25 $\tilde{\tilde{g}}_{HH}^2$ & 1    & 46.82 & 
                        $D^{*+}K^0$      & 3708.12 $\tilde{\tilde{g}}_{HH}^2$ & 1    & 40.09     \\
                        & $D^{*0}\pi^0$  & 1977.11 $\tilde{\tilde{g}}_{HH}^2$ & 0.50 & 23.59 & 
                        $D^{*0}K^+$      & 3753.87 $\tilde{\tilde{g}}_{HH}^2$ & 1.01 & 40.58     \\
                        & $D^{*0}\eta $  & 418.09 $\tilde{\tilde{g}}_{HH}^2$  & 0.11 & 4.99  & 
                        $D^{*+}_s\eta$   & 1787.63 $\tilde{\tilde{g}}_{HH}^2$ & 0.48 & 19.33     \\
                        & $D^{*+}_s K^-$ & 2060.77 $\tilde{\tilde{g}}_{HH}^2$ & 0.52 & 24.59 & 
                        $D^{*+}_s\pi^0$  & 0.21 $\tilde{\tilde{g}}_{HH}^2$    & -    & -         \\
                        & Total          & 8379.22 $\tilde{\tilde{g}}_{HH}^2$ &      &       & 
                        Total            & 9249.82 $\tilde{\tilde{g}}_{HH}^2$ &      &           \\ \midrule
$3S_{\frac{1}{2}}1^-$   & $D^{*+}\pi^-$  & 2858.80 $\tilde{\tilde{g}}_{HH}^2$ & 1    & 28.03 & 
                        $D^{*+}K^0$      & 2693.33 $\tilde{\tilde{g}}_{HH}^2$ & 1    & 24.31     \\
                        & $D^{*0}\pi^0$  & 1439.96 $\tilde{\tilde{g}}_{HH}^2$ & 0.50 & 14.12 & 
                        $D^{*0}K^+$      & 2724.39 $\tilde{\tilde{g}}_{HH}^2$ & 1.01 & 24.60     \\
                        & $D^{*0}\eta $  & 316.40 $\tilde{\tilde{g}}_{HH}^2$  & 0.11 & 3.10  & 
                        $D^{*+}_s\eta$   & 1327.96 $\tilde{\tilde{g}}_{HH}^2$ & 0.49 & 11.99     \\
                        & $D^{*+}_s K^-$ & 1584.76 $\tilde{\tilde{g}}_{HH}^2$ & 0.55 & 15.54 & 
                        $D^{*+}_s\pi^0$  & 0.15 $\tilde{\tilde{g}}_{HH}^2$    & -    & -         \\
                        & $D^{+}\pi^-$   & 1762.48 $\tilde{\tilde{g}}_{HH}^2$ & 0.62 & 17.28 & 
                        $D^{+}K^0$       & 1701.62 $\tilde{\tilde{g}}_{HH}^2$ & 0.63 & 15.36     \\
                        & $D^{0}\pi^0$   & 887.94 $\tilde{\tilde{g}}_{HH}^2$  & 0.31 & 8.71  & 
                        $D^{0}K^+$       & 1720.12 $\tilde{\tilde{g}}_{HH}^2$ & 0.64 & 15.53     \\
                        & $D^{0}\eta $   & 214.69 $\tilde{\tilde{g}}_{HH}^2$  & 0.07 & 2.10  & 
                        $D^{+}_s\eta$    & 909.34 $\tilde{\tilde{g}}_{HH}^2$  & 0.34 & 8.21      \\
                        & $D^{+}_s K^-$  & 1133.57 $\tilde{\tilde{g}}_{HH}^2$ & 0.40 & 11.11 & 
                        $D^{+}_s\pi^0$   & 0.09 $\tilde{\tilde{g}}_{HH}^2$    & -    & -         \\
                        & Total          & 10198.6 $\tilde{\tilde{g}}_{HH}^2$ &      &       & 
                        Total            & 11077 $\tilde{\tilde{g}}_{HH}^2$   &      &           \\ \midrule
$2D_{\frac{3}{2}}1^-$   & $D^{*+}\pi^-$  & 2330.58 $\tilde{g}_{XH}^2$         & 1    & 11.47 & 
                        $D^{*+}K^0$      & 2786.21 $\tilde{g}_{XH}^2$         & 1    & 10.47     \\
                        & $D^{*0}\pi^0$  & 1177.12 $\tilde{g}_{XH}^2$         & 0.50 & 5.79  & 
                        $D^{*0}K^+$      & 2820.01 $\tilde{g}_{XH}^2$         & 1.01 & 10.60     \\
                        & $D^{*0}\eta $  & 305.84  $\tilde{g}_{XH}^2$         & 0.13 & 1.50  & 
                        $D^{*+}_s\eta$   & 1313.32 $\tilde{g}_{XH}^2$         & 0.47 & 4.93      \\
                        & $D^{*+}_s K^-$ & 1361.16 $\tilde{g}_{XH}^2$         & 0.58 & 6.70  & 
                        $D^{*+}_s\pi^0$  & 0.13 $\tilde{g}_{XH}^2$            & -    & -         \\
                        & $D^{+}\pi^-$   & 6561.64 $\tilde{g}_{XH}^2$         & 2.81 & 32.28 & 
                        $D^{+}K^0$       & 7805.01 $\tilde{g}_{XH}^2$         & 2.80 & 29.33     \\
                        & $D^{0}\pi^0$   & 3317.96 $\tilde{g}_{XH}^2$         & 1.42 & 16.32 & 
                        $D^{0}K^+$       & 7903.78 $\tilde{g}_{XH}^2$         & 2.84 & 29.70     \\
                        & $D^{0}\eta $   & 917.22 $\tilde{g}_{XH}^2$          & 0.39 & 4.51  & 
                        $D^{+}_s\eta$    & 3984.22 $\tilde{g}_{XH}^2$         & 1.43 & 14.97     \\
                        & $D^{+}_s K^-$  & 4353.44 $\tilde{g}_{XH}^2$         & 1.87 & 21.42 & 
                        $D^{+}_s\pi^0$   & 0.36 $\tilde{g}_{XH}^2$            & -    & -         \\
                        & Total          & 20325.00 $\tilde{g}_{XH}^2$        &      &       & 
                        Total            & 26613.00 $\tilde{g}_{XH}^2$        &      &           \\ \midrule
$2D_{\frac{3}{2}}2^-$   & $D^{*+}\pi^-$  & 7197.39 $\tilde{g}_{XH}^2$         & 1    & 44.93 & 
                        $D^{*+}K^0$      & 8658.34 $\tilde{g}_{XH}^2$         & 1    & 40.22     \\
                        & $D^{*0}\pi^0$  & 3634.83 $\tilde{g}_{XH}^2$         & 0.50 & 22.69 & 
                        $D^{*0}K^+$      & 8761.67 $\tilde{g}_{XH}^2$         & 1.01 & 40.70     \\
                        & $D^{*0}\eta $  & 948.72  $\tilde{g}_{XH}^2$         & 0.13 & 5.92  & 
                        $D^{*+}_s\eta$   & 4104.25 $\tilde{g}_{XH}^2$         & 0.47 & 19.07     \\
                        & $D^{*+}_s K^-$ & 4237.69 $\tilde{g}_{XH}^2$         & 0.59 & 26.45  & 
                        $D^{*+}_s\pi^0$  & 0.39 $\tilde{g}_{XH}^2$            &      &           \\ 
                        & Total          & 16018.60 $\tilde{g}_{XH}^2$        &     &        & 
                        Total            & 21524.70 $\tilde{g}_{XH}^2$        &      &           \\ \midrule
$2D_{\frac{5}{2}}2^-$   & $D^{*+}\pi^-$  & 2757.21 $\tilde{g}_{YH}^2$         & 1   & 52.58  & 
                        $D^{*+}K^0$      & 2700.76 $\tilde{g}_{YH}^2$         & 1    & 41.99    \\
                        & $D^{*0}\pi^0$  & 1401.61 $\tilde{g}_{YH}^2$         & 0.51 & 26.73  & 
                        $D^{*0}K^+$      & 2765.74 $\tilde{g}_{YH}^2$         & 1.02 & 43.00     \\
                        & $D^{*0}\eta $  & 216.58  $\tilde{g}_{YH}^2$         & 0.08 & 4.13  & 
                        $D^{*+}_s\eta$   & 964.56 $\tilde{g}_{YH}^2$          & 0.36 & 15.00     \\
                        & $D^{*+}_s K^-$ & 867.94  $\tilde{g}_{YH}^2$         & 0.31 & 16.55  & 
                        $D^{*+}_s\pi^0$  & 0.15 $\tilde{g}_{YH}^2$            &  -   &  -        \\ 
                        & Total          & 5243.34 $\tilde{g}_{YH}^2$         &     &        & 
                        Total            & 6431.22$\tilde{g}_{YH}^2$          &      &           \\ \midrule
$2D_{\frac{5}{2}}3^-$   & $D^{*+}\pi^-$  & 1619.99 $\tilde{g}_{YH}^2$         & 1   & 26.23  & 
                        $D^{*+}K^0$      & 1582.30 $\tilde{g}_{YH}^2$         & 1    & 18.16     \\
                        & $D^{*0}\pi^0$  & 823.41 $\tilde{g}_{YH}^2$          & 0.51 & 13.33  & 
                        $D^{*0}K^+$      & 1620.09 $\tilde{g}_{YH}^2$         & 1.02 & 18.59     \\
                        & $D^{*0}\eta $  & 128.26  $\tilde{g}_{YH}^2$         & 0.08 & 2.08  & 
                        $D^{*+}_s\eta$   & 567.81 $\tilde{g}_{YH}^2$          & 0.36 & 6.51      \\
                        & $D^{*+}_s K^-$ & 515.81 $\tilde{g}_{YH}^2$          & 0.32 & 8.35  & 
                        $D^{*+}_s\pi^0$  & 0.09 $\tilde{g}_{YH}^2$            & -    & -         \\
                        & $D^{+}\pi^-$   & 2008.96 $\tilde{g}_{YH}^2$         & 1.24 & 28.21  & 
                        $D^{+}K^0$       & 2030.05 $\tilde{g}_{YH}^2$         & 1.28 & 23.30     \\
                        & $D^{0}\pi^0$   & 1022.84 $\tilde{g}_{YH}^2$         & 0.63 & 14.36  & 
                        $D^{0}K^+$       & 2079.23 $\tilde{g}_{YH}^2$         & 1.31 & 23.86     \\
                        & $D^{0}\eta $   & 185.76 $\tilde{g}_{YH}^2$          & 0.11 & 2.61   & 
                        $D^{+}_s\eta$    & 834.49 $\tilde{g}_{YH}^2$          & 0.53 & 9.58      \\
                        & $D^{+}_s K^-$  & 817.20 $\tilde{g}_{YH}^2$          & 0.50 & 11.47  & 
                        $D^{+}_s\pi^0$   & 0.11 $\tilde{g}_{YH}^2$            & -    & -         \\
                        & Total          & 7122.23 $\tilde{g}_{YH}^2$         &      &        & 
                         Total           & 8714.18 $\tilde{g}_{YH}^2$         &      &           \\ \bottomrule
    \end{tabular*}
    \label{tab:3s2dwidth}
\end{table}
\begin{table}[tbh!]
    \centering
    \caption{Strong decay widths of $1F$ states of $D$ and $D_s$ mesons using the masses calculated in the present study.}
    \begin{tabular*}{\textwidth}{@{\extracolsep{\fill}}p{1cm}p{1.2cm}cccp{1.2cm}ccc}\toprule
     States ($nL_{s_l}J^P$) & Decay Mode  & Width                & Ratio & Fraction & Decay Mode& Width                & Ratio & Fraction \\ \midrule \midrule
$1F_{\frac{5}{2}}2^+$   &$D^{*+}\pi^-$      & 489.31 $g_{ZH}^2$     & 1    & 13.29  & 
                        $D^{*+}K^0$         & 558.92 $g_{ZH}^2$     & 1    & 10.87      \\
                        & $D^{*0}\pi^0$     & 249.30 $g_{ZH}^2$     & 0.51 & 6.77  & 
                        $D^{*0}K^+$         & 573.06 $g_{ZH}^2$     & 1.02 & 11.14      \\
                        & $D^{*0}\eta $     & 43.08 $g_{ZH}^2$      & 0.09 & 1.17  & 
                        $D^{*+}_s\eta$      & 184.55 $g_{ZH}^2$     & 0.33 & 3.59       \\
                        & $D^{*+}_s K^-$    & 150.36 $g_{ZH}^2$     & 0.31 & 4.08  & 
                        $D^{*+}_s\pi^0$     & 0.03 $g_{ZH}^2$       & -    & -          \\
                        & $D^{+}\pi^-$      & 1353.57 $g_{ZH}^2$    & 2.77 & 36.78  & 
                        $D^{+}K^0$          & 1582.56 $g_{ZH}^2$    & 2.83 & 30.77      \\
                        & $D^{0}\pi^0$      & 691.00 $g_{ZH}^2$     & 1.41 & 18.78  & 
                        $D^{0}K^+$          & 1623.01 $g_{ZH}^2$    & 2.90 & 31.56      \\
                        & $D^{0}\eta $      & 141.96 $g_{ZH}^2$     & 0.29 & 3.86  & 
                        $D^{+}_s\eta$       & 620.72 $g_{ZH}^2$     & 1.11 & 12.07      \\
                        & $D^{+}_s K^-$     & 561.56 $g_{ZH}^2$     & 1.15 & 15.26  & 
                        $D^{+}_s\pi^0$      & 0.07 $g_{ZH}^2$       & -    & -          \\
                        & Total             & 3680.14 $g_{ZH}^2$    &      &        & 
                        Total               & 5142.92 $g_{ZH}^2$    &      &            \\ \midrule
$1F_{\frac{5}{2}}3^+$   &$D^{*+}\pi^-$      & 1223.27 $g_{ZH}^2$    & 1    & 52.50  & 
                        $D^{*+}K^0$         & 1421.42 $g_{ZH}^2$    & 1    & 42.43      \\
                        & $D^{*0}\pi^0$     & 623.26 $g_{ZH}^2$     & 0.51 & 26.75  & 
                        $D^{*0}K^+$         & 1457.21 $g_{ZH}^2$    & 1.02 & 43.50      \\
                        & $D^{*0}\eta $     & 107.70 $g_{ZH}^2$     & 0.09 & 4.62  & 
                        $D^{*+}_s\eta$      & 471.11 $g_{ZH}^2$     & 0.33 & 14.06      \\
                        & $D^{*+}_s K^-$    & 375.90 $g_{ZH}^2$     & 0.31 & 16.13  & 
                        $D^{*+}_s\pi^0$     & 0.07 $g_{ZH}^2$       &       &           \\
                        & Total             & 2330.14 $g_{ZH}^2$    &       &       & 
                        Total               & 3349.81 $g_{ZH}^2$    &       &           \\ \midrule
$1F_{\frac{7}{2}}3^+$   &$D^{*+}\pi^-$      & 3650.15 $g_{RH}^2$    & 1     & 59.17  & 
                        $D^{*+}K^0$         & 3229.73 $g_{RH}^2$    & 1     & 44.04     \\
                        & $D^{*0}\pi^0$     & 1874.49 $g_{RH}^2$    & 0.51  & 30.39  & 
                        $D^{*0}K^+$         & 3361.45 $g_{RH}^2$    & 1.04  & 45.84     \\
                        & $D^{*0}\eta $     & 162.81 $g_{RH}^2$     & 0.04  & 2.60  & 
                        $D^{*+}_s\eta$      & 741.87 $g_{RH}^2$     & 0.23  & 10.12      \\
                        & $D^{*+}_s K^-$    & 493.07 $g_{RH}^2$     & 0.13 & 7.83  & 
                        $D^{*+}_s\pi^0$     & 0.21 $g_{RH}^2$       &   -   &  -        \\
                        & Total             & 5952.78 $g_{RH}^2$    &       &       & 
                        Total               & 7333.27 $g_{RH}^2$    &       &           \\ \midrule
$1F_{\frac{7}{2}}4^+$   &$D^{*+}\pi^-$      & 2042.31 $g_{RH}^2$    & 1     & 20.36  & 
                        $D^{*+}K^0$         & 1822.92 $g_{RH}^2$    & 1     & 14.70     \\
                        & $D^{*0}\pi^0$     & 1048.77 $g_{RH}^2$    & 0.51  & 10.45  & 
                        $D^{*0}K^+$         & 1897.00 $g_{RH}^2$    & 1.04  & 15.29     \\
                        & $D^{*0}\eta $     & 91.35 $g_{RH}^2$      & 0.04  & 0.91  & 
                        $D^{*+}_s\eta$      & 420.37 $g_{RH}^2$     & 0.23  & 3.39      \\
                        & $D^{*+}_s K^-$    & 276.99 $g_{RH}^2$     & 0.13 & 2.76  & 
                        $D^{*+}_s\pi^0$     & 0.12 $g_{RH}^2$       & -     & -         \\
                        & $D^{+}\pi^-$      & 3656.43 $g_{RH}^2$    & 1.79 & 36.44  & 
                        $D^{+}K^0$          & 3534.62 $g_{RH}^2$    & 1.94  & 28.49     \\
                        & $D^{0}\pi^0$      & 1882.14 $g_{RH}^2$    & 0.92 & 18.76  & 
                        $D^{0}K^+$          & 3676.36 $g_{RH}^2$    & 2.02  & 29.63     \\
                        & $D^{0}\eta $      & 225.65 $g_{RH}^2$     & 0.11 & 2.25  & 
                        $D^{+}_s\eta$       & 1055.12 $g_{RH}^2$    & 0.58  & 8.50      \\
                        & $D^{+}_s K^-$     & 808.21 $g_{RH}^2$     & 0.39 & 8.06  & 
                        $D^{+}_s\pi^0$      & 0.22  $g_{RH}^2$      & -     & -         \\
                        & Total             & 10031.80 $g_{RH}^2$   &     &        & 
                        Total               & 12406.70 $g_{RH}^2$   &       &           \\ \bottomrule
    \end{tabular*}
    \label{tab:1fwidth}
\end{table}
\section{Results and Discussion}\label{sec:results}
The calculated spin-averaged masses using Song and Lin's potential for $D$($c\bar{q}$) and $D_s$($c\Bar{s}$) mesons are listed in Tables \ref{tab:CwDmass} and \ref{tab:CwDsmass}. The masses with hyperfine splittings are also calculated and given for $D$ mesons in Tables \ref{tab:SPwaveDmeson}, \ref{tab:DwaveDmeson} and for $D_s$ mesons in Tables \ref{tab:SPwaveDsmeson}, \ref{tab:DwaveDsmeson}. The masses are compared with other theoretical models, and experiments are in good agreement. The mixing angles for $nP$ states are in the lower range of different models \cite{godfrey2016}, while for $nD$ states are on the higher range. The mixing angles for $nF$ states are very small. The use of parameters $c$ in the potential suggests that the increase in binding energy with radial quantum number $n$ is not sufficient from the potential itself. This can also be said for a parameter $d$ that is increasing the energy with angular quantum number $l$. These terms are also used in other potentials, like in Refs. \cite{Patel_2016} and \cite{ritu_2024}, where similar parameters are used to explain the properties of charmonium and bottomonium. The Song and Lin's potential was successful in explaining the nature of charmonium without these terms, and we have applied it to the heavy-light system with good outcomes. The hyperfine splitting and other results are discussed below.

The hyperfine splitting of $D$ mesons for $S-$wave are $\Delta(1S)=145$ MeV, $\Delta(2S)=62$ MeV, $\Delta(3S)=42$ MeV, $\Delta(4S)=32$ MeV, and $\Delta(5S)=28$ MeV, which are in fair agreement with available experimental \cite{PDG2022} and theoretical \cite{akraipot2021, godfrey2016, Ruhui2022, Kher2017, Ebert2010} splitting, where $\Delta(1S)=M(^3S_1)-M(^1S_0)$. For the strange charm $D_s$ meson, $\Delta(1S)=80$ MeV, $\Delta(2S)=62$ MeV, $\Delta(3S)=42$ MeV, $\Delta(4S)=32$ MeV and $\Delta(5S)=28$ MeV are also in fare accordance with other available masses \cite{PDG2022, godfrey2016, Yang2023, Ruhui2022, Kher2017, Ebert2010}. 

In the non-strange sector, $1S$ mesons $D$, $D^*$ and $1P$ mesons $D^*_0(2300)$, $D_1(2430)$, $D_1(2420)$ and $D_2^*(2460)$ are well established as $1S_{\frac{1}{2}}0^-$, $1S_{\frac{1}{2}}1^-$, $1P_{\frac{1}{2}}0^+$, $1P_{\frac{1}{2}}1^+$, $1P_{\frac{3}{2}}1^+$ and $1P_{\frac{3}{2}}2^+$ states, respectively, given in the PDG \cite{PDG2022}. In strange sector, $1S$ mesons $D_s(1968)$, $D_s^*(2112)$ and $1P$ mesons $D_{s0}^*(2317)$, $D_{s1}^{'}(2460)$ and $D_{s1}(2536)$ are also well established. 

The calculated masses of mesons are used to construct the Regge trajectories in the ($J, M^2$) and ($n_r, M^2$) planes with the following definitions:
\begin{align}
    J=\alpha_{(n)} M^2+\alpha_{(n)}(0) \label{eq:jm}\\
    n_r=\beta_{(L)} M^2+\beta_{(L)}(0) \label{eq:nrm}
\end{align}
The Eq. \eqref{eq:jm} is in the ($J, M^2$) plane and Eq. ($n_r, M^2$) plane. $\alpha$ and $\beta$ are slopes, and $\alpha_0$ and $\beta_0$ are intercepts of the Regge trajectories. Also, $n=1,2,3,4,5$ are the radial quantum number, and $L=S, P, D, F$ represents the orbital excitations. 
\begin{figure*}
\subcaptionbox{$D$ mesons with unnatural parity in ($J,M^2$) plane. \label{fig:p1}}{\includegraphics[width=0.48\linewidth]{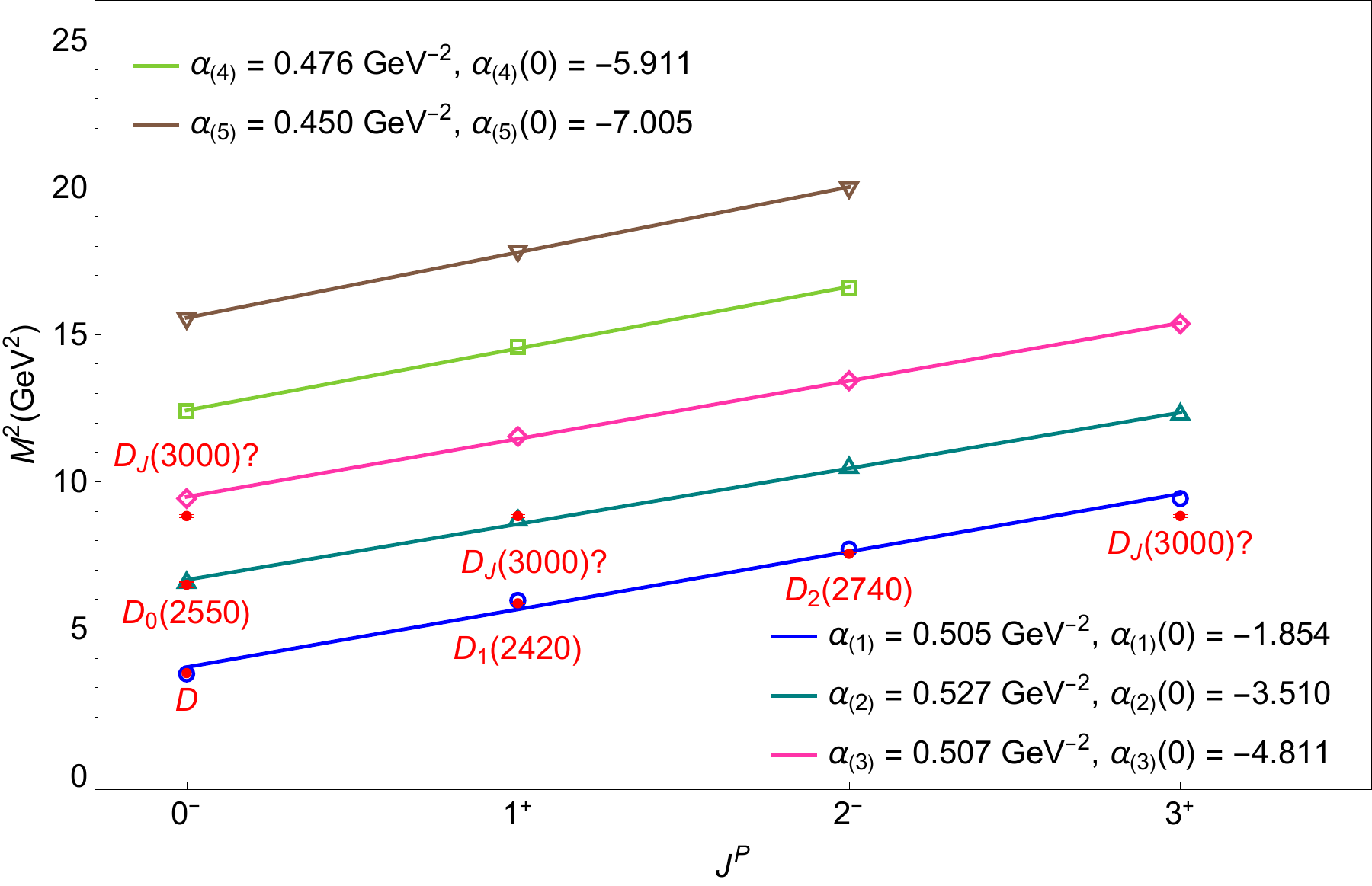}}%
\hfill
\subcaptionbox{$D$ mesons with natural parity in ($J,M^2$) plane. \label{fig:p2}}{\includegraphics[width=0.48\linewidth]{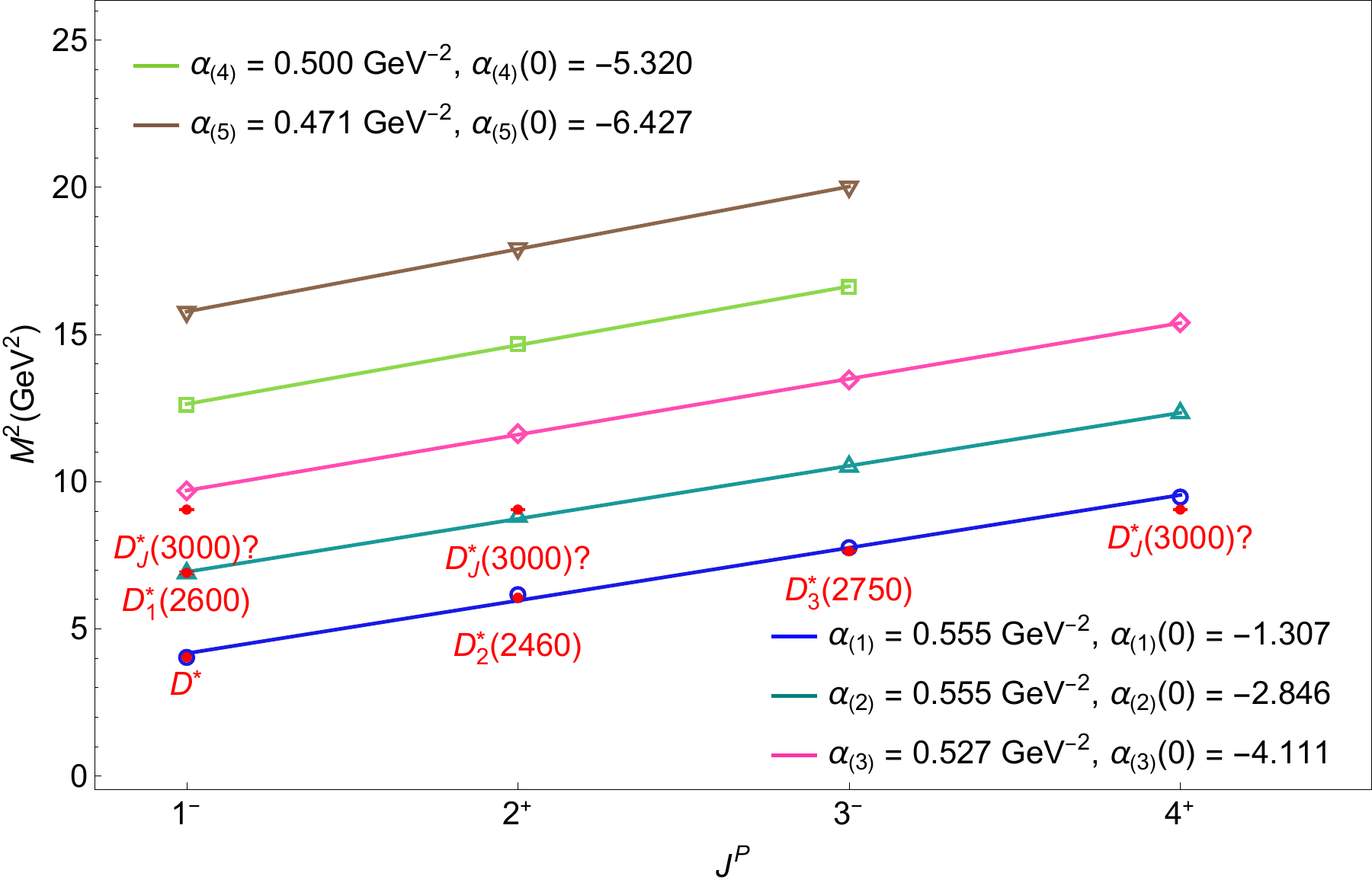}}\\%
\hfill
\subcaptionbox{$D_s$ mesons with unnatural parity in ($J,M^2$) plane. \label{fig:p3}}{\includegraphics[width=0.48\linewidth]{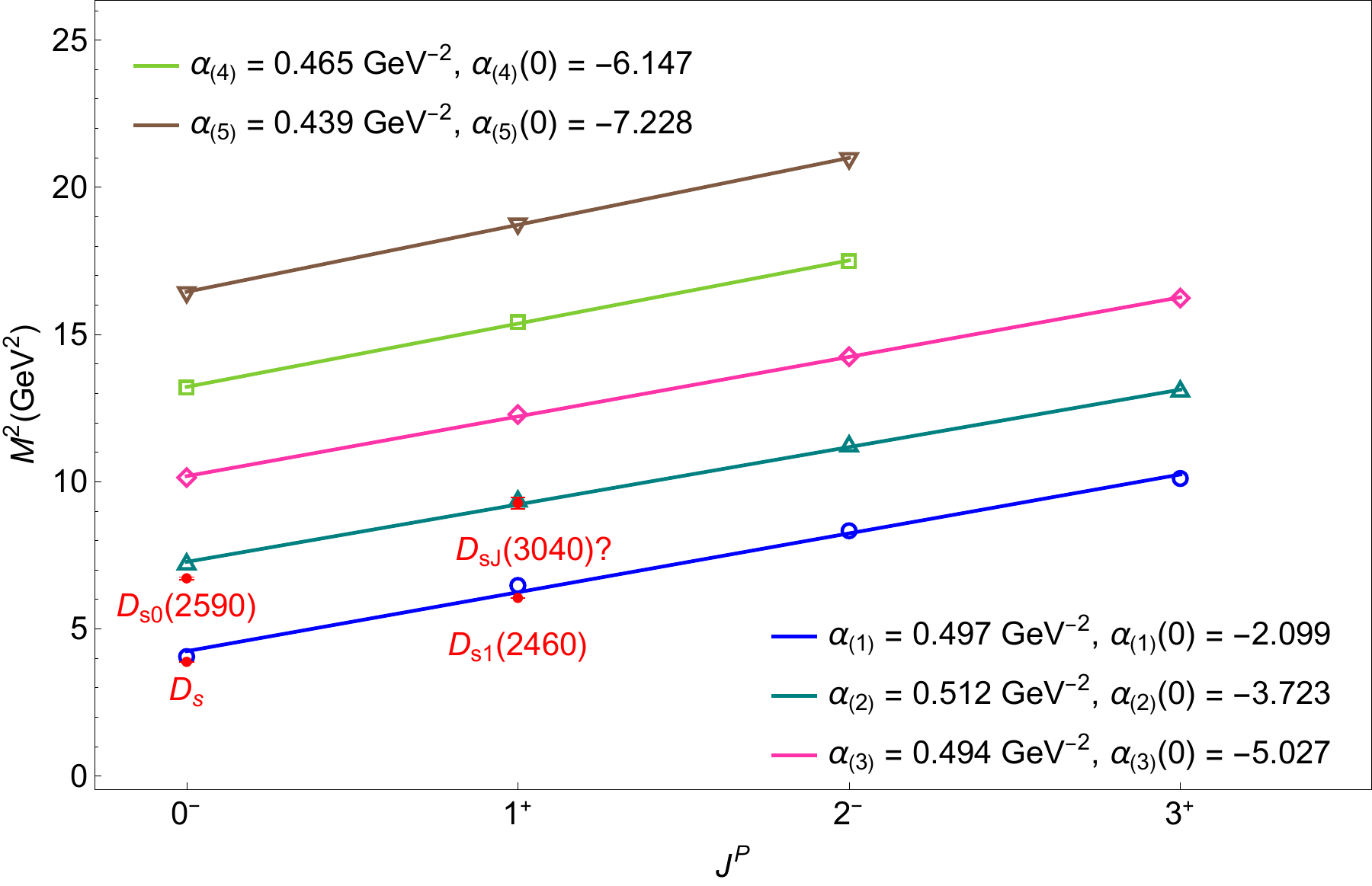}}%
\hfill
\subcaptionbox{$D_s$ mesons with natural parity in ($J,M^2$) plane. \label{fig:p4}}{\includegraphics[width=0.48\linewidth]{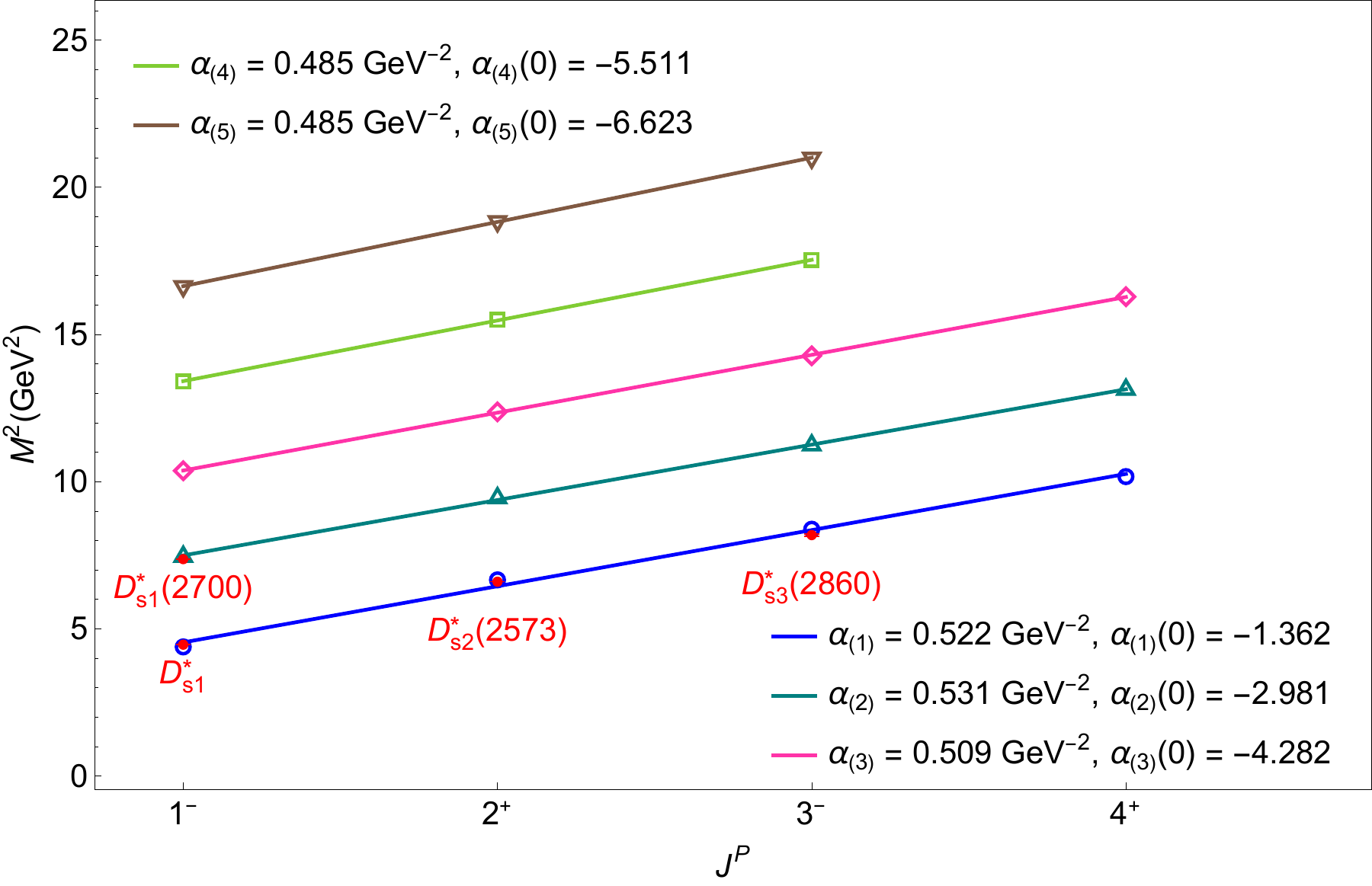}}\\%
\hfill
\subcaptionbox{The spin average masses of $D$ meson in ($n_r,M^2$) plane.  \label{fig:p5}}{\includegraphics[width=0.48\linewidth]{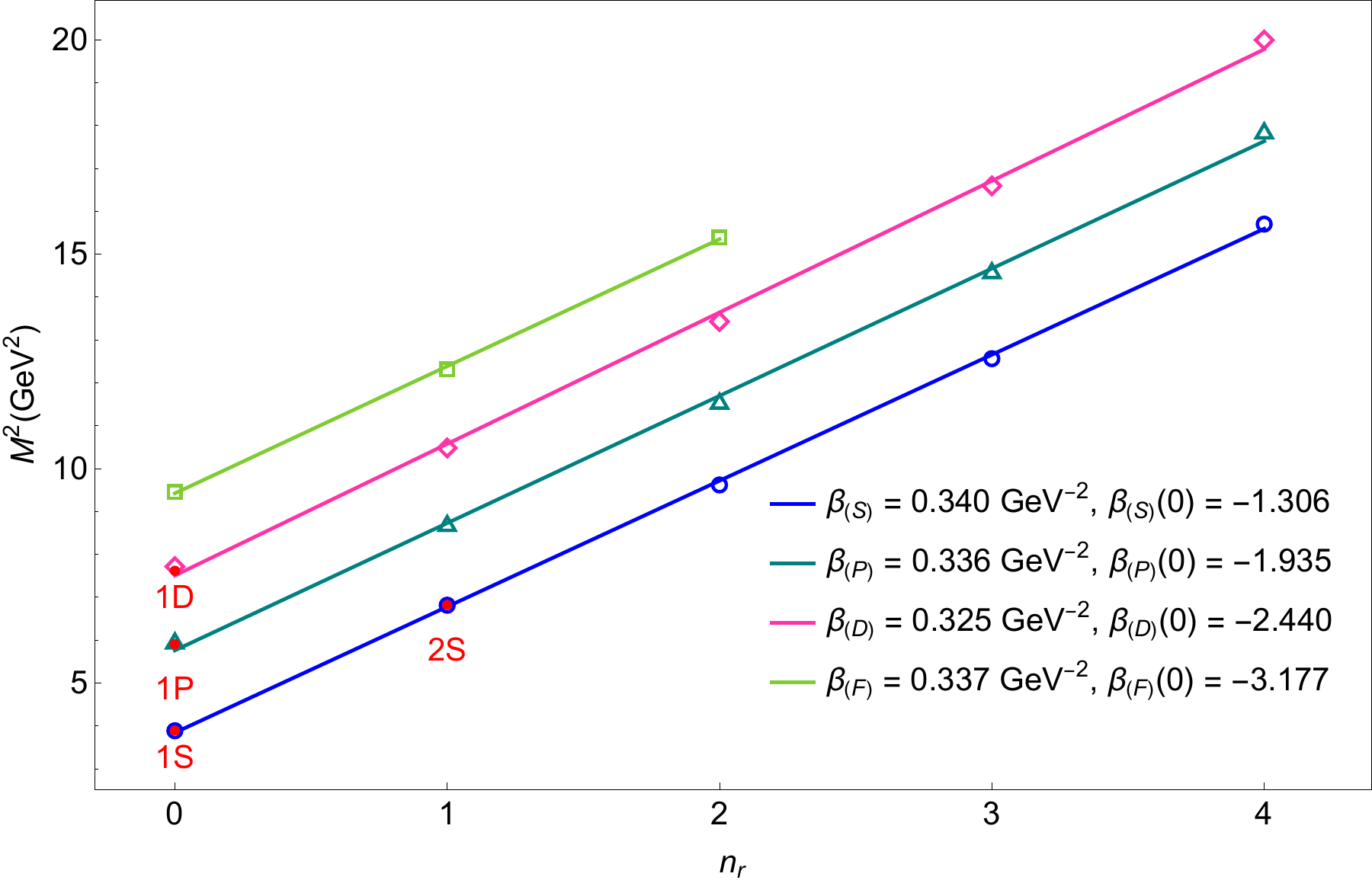}}%
\hfill
\subcaptionbox{The spin average masses of $D_s$ meson in ($n_r,M^2$) plane. \label{fig:p6}}{\includegraphics[width=0.48\linewidth]{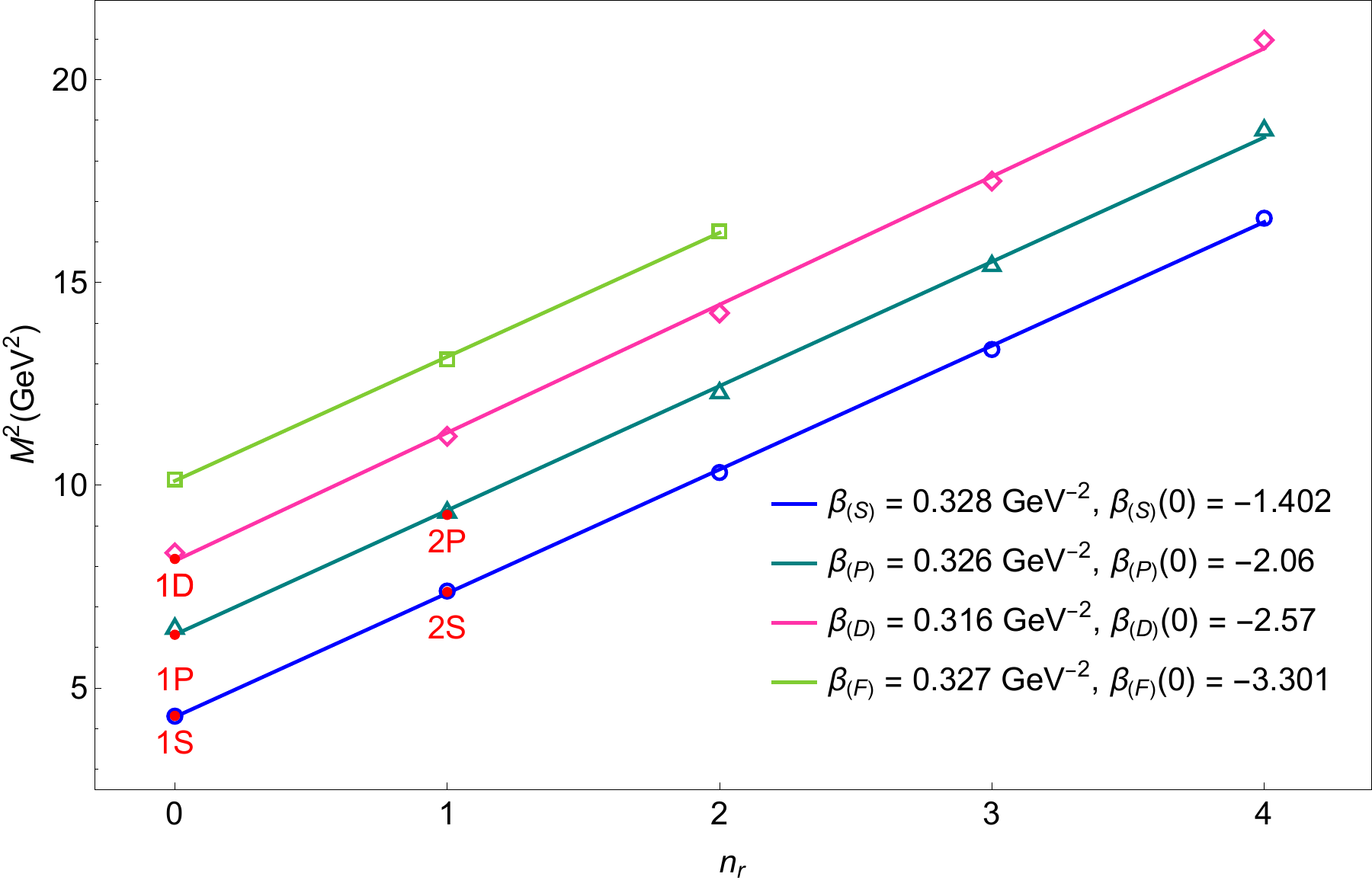}}%
\caption{Regge trajectories for both $D$ and $D_s$ mesons are shown. Hollow markers are used for calculated masses of mesons, and red dots are used for available experimental masses, with their names shown along with the corresponding figures. The Regge trajectories in Figs. \ref{fig:p1}, \ref{fig:p2}, \ref{fig:p3}, \ref{fig:p4} are in the ($J,M^2$) plane and in Figs. \ref{fig:p5}, and \ref{fig:p6} are in the ($n_r,M^2$) plane. The plotted lines are the linear fit on the calculated masses, and the corresponding slopes and intercepts are mentioned in the Regge trajectories above.}
\label{fig:allregge}
\end{figure*}
The Regge trajectories for unnatural ($P=(-1)^{J+1}$) and natural ($P=(-1)^J$) parity states are shown in Figs. \ref{fig:p1} and \ref{fig:p2} for $D$ mesons and in Figs. \ref{fig:p3} and \ref{fig:p4} for $D_s$ mesons are shown. The slopes and intercepts for corresponding Regge trajectories are also shown in Fig. \ref{fig:allregge}. All trajectories are parallel and equidistant to each other, and the calculated masses also fit nicely with the linear trajectories. 

In the following sections, many properties like masses, decay width, partial width ratios, and strong coupling constants are discussed and compared with the available experimental and theoretical values. This discussion and analysis are used to give the assignment to experimentally observed states. Further, some properties that can be observed in the future by experimental facilities are discussed to clear the status of some observed states.
\subsection{\texorpdfstring{$1^3P_2$($1P_{\frac{3}{2}}2^+$)}{TEXT}}
The calculated mass of D($1^3P_2$) state is 2482 MeV, which is close to mass $2461.1 \pm 0.7$ MeV of $D_{2}^*(2460)$ given in PDG \cite{PDG2022}. Using the total decay width $\Gamma(D^*_2(2460))=47.3 \pm 0.8$ MeV, we get $g_{TH}=0.385\pm0.003$, which is in accordance with $g_{TH}$ computed in other Refs. $0.43\pm0.01$\cite{colangelo2012}, $0.40\pm0.003$\cite{Pandya2021}, $0.40\pm0.01$\cite{Pallavi2018}, $0.43\pm0.05$ \cite{wang2014}. We found the ratio
    $\frac{\Gamma(D^{*}_2(2460)\rightarrow D^+\pi^-)}{\Gamma(D^{*}_2(2460)\rightarrow D^{*+}\pi^-)}=2.14$,
which is in agreement with the average of experimental value $1.52\pm0.14$ \cite{PDG2022} and other theoretical estimates $2.29$ \cite{wanghqet2013}, 2.26 \cite{Pandya2021}, 1.96 \cite{songchen2015b}, 1.70 \cite{Ruhui2022}. For $D^{*}_2(2460)$, BaBar collaboration \cite{babar2009b} measured the ratio $\frac{\Gamma(D^+\pi^-)}{\Gamma(D^+\pi^-)+\Gamma(D^{*+}\pi^-)}=0.62\pm0.03\pm0.02$, we have estimated this ratio to be 0.68. In the $D_s$ sector, the calculated mass of $D_s(1P_{\frac{3}{2}}2^+)$ state is 2581 MeV, which is very close to $D_{s2}^*(2573)$ state mass $2569.1\pm0.8$ given in PDG \cite{PDG2022}. Using the calculated coupling $g_{TH}$, we find the total decay width 19.26 MeV is also close to $16.9\pm0.7$ from PDG \cite{PDG2022}. The ratio $\frac{\Gamma(D_{s2}^*(2573)\rightarrow D^{*0}K^+)}{\Gamma(D_{s2}^*(2573)\rightarrow D^0K^+)}=0.11$ is in well accordance with the experimental value $\frac{\Gamma(D_{s2}^*(2573)\rightarrow D^{*0}K^+)}{\Gamma(D_{s2}^*(2573)\rightarrow D^0K^+)}<0.33$ \cite{PDG2022}. $D_2^*(2460)$ and $D_{s2}^*(2573)$ are spin partners of each other, and the analyzed ratios support the given assignment.
\subsection{\texorpdfstring{$1^3D_1$($1D_{\frac{3}{2}}1^-$)}{TEXT}}
The calculated mass of $D(1^3D_1)$ state is 2759 MeV. In Table \ref{tab:1p1d}, the $D\pi$ is the dominant decay mode for this state. The candidate available for this state is $D^*_1(2760)^0$ with mass $2781\pm18\pm13$ MeV and decay width $177\pm32\pm21$ MeV reported by LHCb \cite{Lhcb2015b}. The total decay width of $D^*_1(2760)^0$ estimates the coupling $g_{XH}$ to be $0.243\pm0.026$. The $D^*_1(2760)^0$ is also studied as a mixing of $2^3S_1-1^3D_1$ states by other theoretical models \cite{songchen2015b, Ruhui2022}. In the $D_s$ family, the mass of $D_s(1^3D_{1})$ state is 2868 MeV, which is in good agreement with other theoretical models given in Table \ref{tab:DwaveDsmeson}. The reported $D^*_{s1}(2860)^{\pm}$ state by LHCb \cite{Lhcb2014} with mass $2859\pm12\pm6\pm23$ MeV and decay width $159\pm23\pm 27\pm72$ MeV is a candidate of $D_s(1^3D_{1})$ state. The value of coupling $g_{XH}$ from total decay width $D^*_{s1}(2860)^{\pm}$ we get is $0.194\pm0.022$. The average of coupling $g_{XH}$ estimated from $D^*_1(2760)^0$ and $D^*_{s1}(2860)^{\pm}$ states is $0.219\pm0.017$, which is close to the previous prediction $0.19\pm0.049$ \cite{wang2014}, $0.12$ \cite{ritupallavi2023}, $0.41\pm0.02$ \cite{pallavi2019} from the bottom meson studies.  
\subsection{\texorpdfstring{$1^3D_3$($1D_{\frac{5}{2}}3^-$)}{TEXT}}
The estimated mass of $D(1^3D_3)$ state is 2785 MeV. The $D^*_3(2750)$ resonance can be interpreted as $D(1^3D_3)$ state with observed mass $2763.1\pm3.2$ MeV and decay width $66\pm5$ MeV \cite{PDG2022}. The decay width of $D(1^3D_3)$ from Table \ref{tab:1p1d}, 493.89 $g_{YH}^2$ is used to estimate the coupling $g_{YH}=0.366\pm0.014$. The computed ratio $R_{\pi}(D^*_3(2750))=1.88$ for $D^*_3(2750)$ is in agreement with other HQET studies like \cite{Pandya2021} but not in agreement with other models like $R_{\pi}(D^*_3(2750))=1.1$ \cite{Ruhui2022}. The ratio $R_{\pi}$ is defined as
\begin{align}
    R_{\pi}(D(1^3D_3))=\frac{\Gamma(D(1^3D_3)\rightarrow D\pi)}{\Gamma(D(1^3D_3)\rightarrow D^{*}\pi)}
\end{align}
Further experimental observations are needed to understand the nature of $D^*_3(2750)$. In the strange sector, the mass computed for $D_s(1^3D_3)$ state is 2895 MeV. The $D^*_{s3}(2860)^{\pm}$ has a mass $2860.5\pm2.6\pm6.5$ MeV which makes it a candidate for the $D_s(1^3D_3)$ state. After putting the value $g_{YH}$ the decay width of $D_s(1^3D_3)$ state is estimated to be $66\pm5$ MeV, which is close to the experimental value $53\pm7\pm7$ MeV. We computed the ratio $R_K(D^*_{s3}(2860))$ given as,
\begin{align}
    R_K(D^*_{s3}(2860))=\frac{\Gamma(D^*_{s3}(2860))\rightarrow D^{*}K)}{\Gamma(D^*_{s3}(2860))\rightarrow DK)}=0.43
\end{align}
which suggests that $DK$ channel is more dominant than $D^*K$. The experimental value of the ratio $R_K(D^*_{s3}(2860))=1.04\pm0.17\pm0.20$ \cite{PDG2022}, which is higher than our result. The results of the other calculations of $R_K(D^*_{s3}(2860))=0.73$ \cite{Ruhui2022} and $R(K)=0.802$ \cite{songchen2015a} also suggest that further experimental observation of $D^*_{s3}(2860)^{\pm}$ with the ratio $R_K(D^*_{s3}(2860))$ is needed to clarify the behaviour. By above agreement in mass and total decay width of $D_{s3}^*(2860)$, we tentatively assign it as $1^3D_3$ states of $Y-$ field in HQET.

\subsection{\texorpdfstring{$1D_2$($1D_{\frac{3}{2}}2^-$)}{TEXT} and \texorpdfstring{$1D^{'}_2$($1D_{\frac{5}{2}}2^-$)}{TEXT}}
The masses of $D(1D_2)$ and $D(1D^{'}_2)$ are 2773 MeV and 2776 MeV, respectively, given in Table \ref{tab:DwaveDmeson}, which are the physical states from the mixing of states $1^1D_2$ and $1^3D_2$ as discussed above. From the above-computed values of $g_{XH}$ and $g_{YH}$, the decay widths of $1D_2$ and $1D^{'}_2$ are found to be $110\pm23$ MeV and $35.7\pm2.7$ MeV respectively. The $D_2(2740)^0$ may be a candidate for $D(1^1D_2)$ or $D(1D^{'}_2)$ state with mass 2747$\pm6$ MeV, and decay width $88\pm19$ MeV \cite{PDG2022}. 
From Table \ref{tab:1p1d}, the $1D^{'}_2$ state is the compact state which is also supported by \cite{Ruhui2022, songchen2015b, Batra2015, Pandya2021, godfrey2016, tan2022}. Taking $D_2(2740)^0$ as the $1D^{'}_2$ state, we get the coupling $g_{YH}=0.57\pm0.06$. Using the earlier value of $g_{YH}$, we get an average of $g_{YH}=0.468\pm0.031$ is close to the estimated values $0.42$ \cite{colangelo2012}, 0.70 \cite{Batra2015}, $0.49\pm0.039$ \cite{Pandya2021}, $0.61\pm0.05$ \cite{Pallavi2018}.
In the strange sector, $D_s(1D_{2})$ and $D_s(1D^{'}_{2})$ states have masses 2881 MeV and 2886 MeV, and decay widths estimated at $126\pm20$ MeV and $50\pm7$ MeV. The experimental observation of these states in strange sector is currently missing, which we hope to see in the future. 
\subsection{\texorpdfstring{$2^1S_0$($2S_{\frac{1}{2}}0^-$)}{TEXT} and \texorpdfstring{$2^3S_1(2S_{\frac{1}{2}}1^-)$}{TEXT}}
The radially excited states are an essential part of the spectrum of mesons to understand the quark and anti-quark pair dynamics. The $D(2^1S_0)$ and $D(2^3S_1)$ states have calculated mass of 2563 MeV and 2625 MeV, respectively. The experimental candidate for $D(2^1S_0)$ state is $D_0(2550)^0$ with mass $2549\pm19$ MeV and decay width $165\pm24$ MeV \cite{PDG2022}. The candidate for $D(2^3S_1)$ state is $D^*_1(2600)$ with mass $2627\pm10$ MeV and decay width $141\pm23$ MeV. In Table \ref{tab:2s2ppwidth}, the state $2S_{\frac{1}{2}}0^-$ decay in $D^*\pi^-$ with $\mathcal{BR}=66.13\%$ supports the experimental observation of $D_0(2550)^0$ in the $D^*\pi^-$ channel by LHCb collaboration \cite{LHCb2020,LHCb2013} and BaBar collaboration \cite{babar2010}. The coupling $\tilde{g}_{HH}$ is found to be $0.375\pm0.027$, considering the $D_0(2550)^0$ as $D(2^1S_0)$ state. The $D^*_1(2600)$ is reported in both $D\pi$ and $D^*\pi$ channels by LHCb collaboration \cite{LHCb2020, LHCb2016prd, LHCb2013} and BaBar collaboration \cite{babar2010}. From the experimental decay width of $D^*_1(2600)$ and the total width in Table \ref{tab:2s2ppwidth} of $2S_{\frac{1}{2}}1^-$, we get the coupling $\tilde{g}_{HH}=0.262\pm0.021$. The theoretical ratio $R_{\pi}(2^3S_1)$ is higher than the reported value $0.32\pm0.02\pm0.09$ \cite{PDG2022, babar2010}. Many theoretical models have studied the $D^*_1(2600)$ as the state of $2S-1D$ mixing \cite{songchen2015b, hao2020, liye2014, Ruhui2022}. In the $D_s$-meson sector, the computed masses of $D_s(2^1S_0)$ and $D_s(2^3S_1)$ are 2682 MeV and 2730 MeV with splitting ($m(2^3S_1)-m(2^1S_0)$) of 48 MeV is in agreement with the other theoretical models given in Table \ref{tab:SPwaveDsmeson}. The $D_{s1}^*(2700)^{\pm}$ is a candidate for $2^3S_1$ state with mass 2714 MeV and decay width $122\pm10$ MeV.  Using the decay width of $D_{s1}^*(2700)^{\pm}$, the coupling $\tilde{g}_{HH}$ is calculated as $0.255\pm0.010$. The theoretically calculated ratio $R_K(D_{s1}^*(2700))=0.96$ is in excellent agreement with the experimental value $0.91\pm0.13\pm0.12$ reported by BaBar collaboration \cite{babar2009}. 
The average $\tilde{g}_{HH}$ from the above-mentioned values is $0.269\pm0.009$. The total decay width of $D_s(2^1S_0)$ is calculated as $63\pm4$ MeV using the average value of $\Tilde{g}_{HH}$. The newly reported $D_{s0}(2590)^+$ with mass $2591\pm6\pm7$ MeV and decay width $89\pm16\pm12$ MeV is suggested to be a strong candidate for the $D_s(2^1S_0)$ state from LHCb collaboration \cite{LHCb2021s}. Our calculated decay width for this state is close to the lower limits of the experimental value. We tentatively assign $D_{s0}(2590)^+$ as the candidate for $D_s(2^1S_0)$ state. The $D_{s0}(2590)^+$ is observed to be decaying in the $D^+K^+\pi^-$ final state. Further experiment observations may shed some light on the other decay channels and ratios of the $D_{s0}(2590)^+$.

\subsection{\texorpdfstring{$D_J(3000)$}{TEXT}, \texorpdfstring{$D_J^{*}(3000)$}{TEXT} and \texorpdfstring{$D_{2}^{*}(3000)$}{TEXT}}
The $D^*_J(3000)^{0}$ and $D_J(3000)^0$ states were observed by LHCb collaboration in 2013 \cite{LHCb2013}. The $D^*_J(3000)^{0}$ is reported to be decaying in the $D\pi$ channel with mass $3008.1\pm4.0$ MeV and decay width of $110.5\pm11.5$ MeV. The $D_J(3000)^0$ is observed in the $D^{*+}\pi^-$ channel with mass $2971.8\pm8.7$ MeV and decay width $188.1\pm44.8$ MeV. 
As LHCb collaboration \cite{LHCb2013} recommended, the $D_J(3000)^0$ with unnatural parity and $D^*_J(3000)$ with natural parity. The suitable states for $D_J(3000)^0$ in the mass range are $3^1S_0(3S_{\frac{1}{2}}0^-)$, $2P_1(2P_{\frac{1}{2}}1^+)$, $2P^{'}_1(2P_{\frac{3}{2}}1^+)$, $1F_3(1F_{\frac{5}{2}}3^+)$, $1F^{'}_3(1F_{\frac{7}{2}}3^+)$. While, the $D^*_J(3000)^{0}$ resonance is analysed as $3^3S_1(3S_{\frac{1}{2}}1^-)$, $2^3P_0(2P_{\frac{1}{2}}0^+)$, $2^3P_2(2P_{\frac{3}{2}}2^+)$, $1^3F_2(1F_{\frac{5}{2}}2^+)$, and $1^3F_4(1F_{\frac{7}{2}}4^+)$.

The masses of $3^1S_0$ and $3^3S_1$ are 3070 MeV and 3112 MeV, respectively. The mass difference $\Delta(3S-2S)$ between states $3^1S_0$ and $2^1S_0$ is about 500 MeV, consistent with other models. Considering both experimental resonances $D_J^*(3000)$ and $D_J(3000)$ as spin partners, the splitting $\Delta(3S)=m(3^3S_1)-m(3^1S_0)=42$ MeV is close to the mass splitting of $36\pm10$ MeV between the experimental masses of $D_J^*(3000)$ and $D_J(3000)$. But the calculated masses of $3^1S_0$ and $3^3S_1$ states are about 100 MeV higher than the measured masses of $D_J^*(3000)$ and $D_J(3000)$. Also, the ratio $R_{\pi}(3^3S_1)=0.62$ suggests that $3^3S_1$ state is in conflict with the observation of $D^*_J(3000)$ is the $D\pi$ channel. Considering $D_J(3000)$ state as a candidate of $3^1S_0$ state, the coupling $\tilde{\tilde{g}}_{HH} = 0.150\pm0.018$ is estimated.

The masses of  $2^3P_0$, $2P_1$, $2P^{'}_1$ and $2^3P_2$ are 2881 MeV, 2947 MeV, 2956 MeV, and 2968 MeV, respectively. The $2^3P_0$ state decay width given in Table \ref{tab:2s2ppwidth}, supports the assignment of $D^*_J(3000)$ as $2^3P_0$. The branching fraction of both $D^+\pi^-$ and $D^0\pi^0$ channels are more than $60\%$ and using the total decay width, we can estimate the coupling $\tilde{g}_{SH}$ to be $0.115\pm0.006$. Analyzing the $D^*\pi$ channel for the $D_J(3000)$, we can assign it as $2P_1$ state, and the coupling $\tilde{g}_{SH}$ using the total decay width is $0.162\pm0.019$.  Among the $2P$ states, the $2^3P_0$ and $2P_1$ states are broad, and $2P^{'}_1$ and $2^3P_2$ are the narrow states. The $D_J(3000)$ may also be $2P^{'}_1$ state, and the coupling $\tilde{g}_{TH}$ is calculated to be $0.205\pm0.024$ by using the total decay widths. Also, the $D^*_J(3000)$ state can be considered as the $2^3P_2$ state, and this gives the value of the coupling $\tilde{g}_{TH}=0.134\pm0.007$. The average of this coupling is $0.139\pm0.007$. The ratio $R_{\pi}(2^3P_2)=1.08$ may be a good indication of reported $D\pi$ channel for $D_J^*(3000)$ to be $2^3P_2$ state. If the $D_J^*(3000)$ is $2^3P_2$ state then we propose that $D_J^*(3000)$ should be searched in $D^*\pi$ channel. Also, the branching ratio $BR(D_s K)$ is about 30$\%$ for both $2^3P_0$ and $2P_1$ whereas the branching ratio $BR(D_s K)$ is only about 15$\%$ for both $2P^{'}_1$ and $2^3P_2$. This can also be verified by future experiments.

The $1F$ states are also candidates for both $D_J(3000)$ and $D^*_J(3000)$. 
The calculated masses of $1F$ states are  3070 MeV, 3070 MeV, 3077 MeV, and 3078 MeV of $1^3F_2$, $1F_3$, $1F^{'}_3$, and $1^3F_4$ states, respectively. The $1^3F_2$ and $1^3F_4$ states have a dominant $D\pi$ decay channel, and the ratio $R_{\pi}(1^3F_2)=2.77$ is greater than the $R_{\pi}(1^3F_4)=1.79$. Thus, the $D^*_J(3000)^{0}$ is tentatively preferred to be $1^3F_2$ state, and using its experimental decay width, we calculate the coupling $g_{ZH}=0.173\pm0.009$. The possibility of $D^*_J(3000)^{0}$ being $1^3F_4$ state may not be ignored. The $1F_3$ and $1F^{'}_3$ states are possible candidates of $D_J(3000)$ as they have $D^*\pi$ decay channels with branching ratio $BR(D^*\pi)\approx 80\%$. Using the total decay widths, we estimate the coupling $g_{ZH}=0.284\pm0.034$ and the coupling $g_{RH}=0.178\pm0.021$. The branching fraction $BR(D_s K)$ for $1^3F_2$ and $1F_3$ are 15$\%$. The branching fraction $BR(D_s K)$ for $1F^{'}_3$ and $1^3F_4$ states are about 8$\%$. These fractions may be observed in future experiments to clear the properties of both $D_J(3000)$ and $D^*_J(3000)$.

In the year 2016, LHCb collaboration \cite{LHCb2016prd} reported $D^*_2(3000)^0$ with mass $3214\pm29$ MeV and decay width $186\pm38$ MeV. The LHCb proposed $D^*_2(3000)^0$ to be a different state than $D^*(3000)$ reported previously \cite{LHCb2013}, but it may not be ruled out that both states are the same. The mass and decay width of $D^*_2(3000)^0$ are higher than the $D^*(3000)$ resonance. The preferred assignments suggested for $D^*_2(3000)^0$ resonance by LHCb collaboration are of $2P$ or $1F$ family. The masses for these states are already discussed above, and treating the $D^*_2(3000)^0$ and $D^*(3000)^0$ as different resonances, we assign the $D^*_2(3000)^0$ as a $1^3F_2$ state. This assignment is favored as the mass of the $1^3F_2$  state is higher than the mass of the $2^3P_2$ state. Using the total decay width of $D^*_2(3000)^0$ and the $1^3F_2$ state from the table \ref{tab:1fwidth}, we compute the coupling $g_{ZH}$ to be $0.225\pm0.023$. In the view of $D^*_2(3000)^0$ assigned as $1^3F_2$, we prefer the $D^*(3000)^0$ as a $2P$ state. To distinguish between the $1^3F_2$ and $2^3P_2$ states, we estimate ratio $\frac{\Gamma(D_sK)}{\Gamma(D^*_sK)}$ for  $1^3F_2$ state to be 3.73 and for $2^3P_2$ state to be 1.60. Also, the ratio $R_{\pi}(1^3F_2)\approx 2.5$ and $R_{\pi}(2^3P_2)\approx1$ may be helpful in distinguishing between states in the experimental reports. 

We tentatively assign $D^*_2(3000)^0$ as $1^3F_2$ state on the basis of masses and decay ratio $R_{\pi}$. Considering $D^*_J(3000)$ and $D_J(3000)$ as spin partners, we tentatively assign them as $2P(0^+,1^+)$ states with the coupling $\tilde{g}_{SH}$ estimated to be $0.115\pm0.006$. If $D^*_J(3000)$ and $D_J(3000)$ resonances are not spin partners, then the assignment of $D_J(3000)$ as $3S(0^-)$ state can be favored with coupling $\tilde{\tilde{g}}_{HH}=0.150\pm0.018$.

\subsection{\texorpdfstring{$D_{sJ}(3040)$}{TEXT}}
The BaBar collaboration \cite{babar2009} in 2009 reported a resonance $D_{sJ}(3040)^{\pm}$ in the inclusive production of $D^*K$ in the $e^+e^-$ annihilation with mass $3044\pm8^{+30}_{-5}$ MeV and decay width of $239\pm35^{+46}_{-42}$ MeV. The  $D_{sJ}(3040)$ is a candidate for $2P_1$ and $2P^{'}_1$ states. The masses of these states are 3054 MeV and 3062 MeV, given in Table \ref{tab:SPwaveDsmeson}, which looks close to the measured mass. The decay widths of $2P$ states are given in Table \ Ref {tab:2s2ppwidth} with the possible channels. The decay width of $2P_1$ state is broader than the $2P^{'}_1$ state. Considering the $D_{sJ}(3040)$ as $2P_1$ state and using the total decay widths, we estimate the $\tilde{g}_{SH}=0.159\pm0.012$. The average from all the calculated values of $\tilde{g}_{SH}$ is $0.127\pm0.005$ is in accordance with other estimations $0.12\pm0.03$ \cite{Pallavi2018}, $0.10\pm0.015$ \cite{Pandya2021}. The coupling $\tilde{g}_{SH}$ from the non-strange $2P$ states and their strange partner are in very good agreement with other theoretical calculations, as shown above. We assign the $D_{sJ}(3040)$ resonance as the $D_s(2P_1)$ state of $S-$ field. Table \ref{tab:allcoupling} shows the couplings calculated in the present study. 
\begin{table}[tbh!]
    \centering
    \caption{The estimated couplings for the strong decays of charm and charm-strange mesons.}
    \begin{tabular}{cccccc} \toprule
       Couplings  & $g_{TH}$ &  $g_{XH}$ & $g_{YH}$ & $\tilde{g}_{HH}$ & $\tilde{g}_{SH}$ \\ \midrule
        Values & $0.385\pm0.003$ & $0.219\pm0.017$ & $0.468\pm0.031$ & $0.269\pm0.009$ & $0.115\pm0.006$ \\ \bottomrule
    \end{tabular}
    \label{tab:allcoupling}
\end{table}
\section{Conclusions}\label{sec:conclusion}
We have analyzed the heavy-light charm mesons in the potential of Song and Lin with calculations and results of masses and decay widths in the above sections. The $1S$, $1P$, and $2S$ masses of charm mesons are fitted to estimate the potential parameters. The spectrum of the charm meson is given in Tables \ref{tab:SPwaveDmeson}, \ref{tab:DwaveDmeson}, \ref{tab:SPwaveDsmeson}, and \ref{tab:DwaveDsmeson}. This potential was proposed by modifying the vector term from the leptonic decay width of vector mesons. The application of Song and Lin's potential to the heavy-light system studied in the present study can improve the understanding of quark-antiquark confining. In the present analysis, terms dependent on radial quantum number ($n$) and orbital quantum number ($l$) are taken. The charm quark mass ($m_c$) estimated in the present study is on the higher side. This characteristic of the present potential will also be explored for bottom mesons. The masses obtained for different states are compared with the results of other theoretical models and experimental masses. The hyperfine splittings are computed with the mixing angles for the corresponding states. The calculated masses with variational parameter $\mu$ are in good agreement with other models. Regge trajectories are plotted in the ($J, M^2$) and ($n_r, M^2$) planes are parallel and equidistant. The strong decay widths are analyzed in the framework of heavy quark effective theory (HQET). The fields of the doublet states of mesons with interacting Lagrangians, as well as the final decay width formulae, are given. The total decay widths of different states with the possible channels are computed in terms of the corresponding couplings and shown in the tables \ref{tab:1p1d}, \ref{tab:2s2ppwidth}, \ref{tab:3s2dwidth}, and \ref{tab:1fwidth}. The assignments of observed resonances in different experimental facilities like LHCb and BaBar are done. The total decay widths of observed states are used to estimate the strong couplings. The branching ratios are also estimated and compared with available theoretical and experimental results. The tentative assignments of states are also discussed in the present study. We hope for further observations of the resonances like $D_2(2740)$, $D_J(3000)$, $D^*_J(3000)$, $D_2^*(3000)$, and $D_{sJ}(3040)$ in different decay channels and properties.
\section{Acknowledgment}
The authors gratefully acknowledge the financial support by the
Department of Science and Technology \\(SERB/F/9119/2020), New
Delhi and for Senior Research Fellowship (09/0677(11306)/2021-EMR-I) by Council of Scientific and Industrial Research, New Delhi.

\bibliography{ref}
\bibliographystyle{epj}
\end{document}